\documentclass[aps, prl, reprint, longbibliography, nofootinbib,superscriptaddress,floatfix]{revtex4-2}

\pdfoutput=1

\usepackage{graphicx} 
\usepackage{xcolor}
\usepackage{babel}
\usepackage{amsthm,amsmath,amssymb,mathrsfs,amsfonts}
\usepackage{bbm}
\usepackage{slashed}
\usepackage{verbatim}
\usepackage[T1]{fontenc}
\usepackage[utf8]{inputenc}
\usepackage[colorlinks=true,linktocpage=true,linkcolor=blue,citecolor=blue]{hyperref}
\usepackage{mathbbol}
\usepackage[normalem]{ulem}
\usepackage{svg}
\usepackage{lipsum}
\usepackage{physics}
\usepackage{dcolumn}
\usepackage{bm}
\usepackage{tensor}
\usepackage{comment}
\usepackage{color,overpic,mathtools}
\usepackage{braket,setspace}
\usepackage{cancel}
\usepackage{float}
\usepackage{fancyhdr}
\fancypagestyle{plain}

\newcommand{\PRLsection}[1]{\noindent{\bf\emph{#1.---}}}

\newcommand{\br}[1]{\left(#1\right)}

\renewcommand{\a}{\alpha}
\renewcommand{\b}{\beta}
\renewcommand{\d}{\delta}
\newcommand{\e}{\epsilon}

\newcommand{\m}{\mu}
\newcommand{\n}{\nu}
\renewcommand{\r}{\rho}
\newcommand{\s}{\sigma}
\renewcommand{\l}{\lambda}
\newcommand{\p}{\pi}

\newcommand{\g}{\gamma}

\newcommand{\w}{\omega}

\newcommand{\D}{\Delta}

\newcommand{\mat}{\mathcal}

\begin{document}

\title{Modelling self-consistently beyond General Relativity}

\author{Ramiro Cayuso}
\affiliation{Perimeter Institute for Theoretical Physics, 31 Caroline Street North, Waterloo, ON N2L 2Y5, Canada}
\affiliation{Department of Physics and Astronomy, University of Waterloo, Waterloo, Ontario, N2L 3G1, Canada}
\author{Pau Figueras}
\affiliation{School of Mathematical Sciences, Queen Mary University of London, Mile End Road, London, E1 4NS, United Kingdom}
\author{Tiago França} 
\affiliation{School of Mathematical Sciences, Queen Mary University of London, Mile End Road, London, E1 4NS, United Kingdom}
\author{Luis Lehner}
\affiliation{Perimeter Institute for Theoretical Physics, 31 Caroline Street North, Waterloo, ON N2L 2Y5, Canada}

\begin{abstract}
The majority of extensions to General Relativity display
mathematical pathologies --higher derivatives, character change in equations that
can be classified within PDE theory, and even unclassifiable ones-- that cause severe difficulties to study
them, especially in dynamical regimes. We present here an approach
that enables their consistent treatment and extraction of physical consequences. 
We illustrate this method in the context of single and merging black holes in a highly
challenging beyond GR theory. 
\end{abstract}

\maketitle

\PRLsection{Introduction}
The gravitational wave window provides 
exciting opportunities to further test General Relativity (GR), e.g.,~\cite{LIGOScientific:2021sio}.
Especially in the context of compact binary mergers, gravitational waves
produced by the strongest gravitational fields in
highly dynamical settings arguably represent the best regime to 
explore deviations from GR, e.g.,~\cite{Will:2018bme}.

Such effort, the ability to extract consequences and propel theory forward,
rely on at least having some understanding of the characteristics
of potential departures, to search and 
interpret outcomes~\cite{Flanagan:1997sx}.

Unfortunately, the majority of proposed beyond GR theories have, at a formal level, 
mathematical pathologies which makes their understanding in general scenarios difficult.\footnote{For
a rather broad sample of beyond GR proposals, discussed in the context of cosmology, see~\cite{Clifton:2011jh}.} Such pathologies may include loss of uniqueness, a dynamical change of character in the equations of motion (e.g., from hyperbolic to elliptic) or, even worse, having equations of motion (EOMs) of unknown mathematical type (e.g.,~\cite{Delsate:2014hba,Papallo:2017qvl,Ripley:2019hxt,Bernard:2019fjb,Okounkova:2019dfo,Figueras:2020dzx,Gerhardinger:2022bcw}). 
This, combined with the need to use computational simulations to study the (non-linear/dynamical) regime
of interest, pose unique challenges.
Of note is that the standard mathematical approach to analyse PDEs ~\cite{1989iii} --where
the high-frequency limit is examined-- cannot be applied as it is in such regime
that the problems alluded above arise. Further, such regime is 
incompatible with the very assumptions made to formulate 
most GR extensions which rely on Effective Field Theory (EFT) arguments~\cite{Burgess:2007pt}. Faced with this problem, solid novel ideas must be pursued to understand potential solutions.

We report here on a technique to {\em fix} the underlying
equations of motion to an extent to which the viability of a given
theory can be assessed.\footnote{This approach is motivated in part by the Israel-Stewart formalism for viscous relativistic hydrodynamics~\cite{1979AnPhy.118..341I}, though in such case, higher order corrections are known and can be called for to motivate the strategy.} In particular,
it allows exploring relevant theories within their regime of
validity and, in particular, monitor whether the dynamics keeps
the solution within it for cases of interest.
This technique, 
partially explored in toy models~\cite{Cayuso:2017iqc,Allwright:2018rut} and restricted settings (e.g.,~\cite{Cayuso:2020lca,Franchini:2022ukz}) is here
developed for the general, and demanding scenario, of compact binary mergers. This requires further considerations not arising in the previously simplified regimes.
Specifically, we present the first self-consistent study of both single and binary 
black hole (BH) merger in the context of an EFT of gravity where corrections to GR
come through high powers (naturally argued for)  in the curvature tensor leading to EOMs with
a priori unclassifiable mathematical character.

We adopt the following notation: Greek letters ($\mu$, $\nu$, $\rho$, ...) 
to denote full spacetime indices and Latin letters ($i$, $j$, $k$, ...) 
for the spatial ones. We use the mostly plus metric signature, and set $c=1$.

\PRLsection{Focusing on a specific theory}
While we could take any of a plethora of proposed beyond GR theories --almost all sharing
the problems  alluded to earlier--, for definiteness here we
 consider a specific extension to GR derived naturally
from EFT arguments~\cite{Endlich:2017tqa}. In this approach,
high energy (i.e., above the cutoff scale) degrees of freedom are integrated out, and their effects are
effectively accounted for through higher order operators acting on the lower energy
ones. For the case of gravitational interactions, in vacuum assuming parity symmetry, and accounting for the simplest contribution, 
such an approach yields
under natural assumptions:\footnote{Other
operators at this (and even lower) orders can be considered, though without loss of generality with
regards to our goals we ignore them here so as to not overly complicate the presentation. }
\begin{equation}
    I_\text{eff} = \frac{1}{16\p G}\int d^4x\, \sqrt{-g}\br{R - \frac{1}{\Lambda^6}\, \mat{C}^2 + \cdots}\,,
\end{equation}
where  $\mat{C}=R_{\a\b\g\d}R^{\a\b\g\d}$
and the coupling scale $\Lambda$ has units of $[M_S]^{-1}$ for some scale $M_S$. The equations of motion are $G_{\m\n} = 8\,\e\,H_{\m\n}$, 
with $G_{\m\n}$ the Einstein tensor, $\e \equiv \Lambda^{-6}$ and
\begin{eqnarray}
     H_{\m\n} &= &~\mat{C}\Big[\Box R_{\m\n}-\tfrac{1}{2}\nabla_\m\nabla_\n R-\tfrac{1}{16}\,\mat{C}\,g_{\m\n} -R_{\m\l}\,R^{\l}_{\phantom{\l}\n} \nonumber \\
        & &\hspace{0.5cm}+R^{\a\b}R_{\m\a\n\b}+\tfrac{1}{2}\,R_{\m\s\r\l}R_{\n}^{\phantom{\n}\s\r\l}\Big]  \\
        & &+2(\nabla^\a\mat{C})\big[\nabla_\a R_{\m\n}-\nabla_{(\m}R_{\n)\a}\big]
        +R_{\m\phantom{\a}\n}^{\phantom{\m}\alpha\phantom{\n}\b}\nabla_\a\nabla_\b\mat{C}\,.\nonumber
\end{eqnarray}
$H_{\m\n}$ is covariantly conserved, since it is derived from an action possessing local diffeomorphism invariance.

In GR ($\epsilon=0$) the resulting EOMs can be shown to define a hyperbolic, linearly degenerate, non-linear, second order, PDE system of equations with constraints
(e.g.,~\cite{Sarbach:2012pr}). With suitable coordinate 
conditions, characteristics are given by the light cones and do not cross --thus
shocks or discontinuities cannot arise. The right hand side (RHS) however, spoils all these considerations. Derivative operators higher than
second order appear --which render the equations outside formal PDE classifications.
{\em How is one to approach the study of this problem?} First, one can simplify 
somewhat the EOMs by applying an order reduction  and replace the Ricci tensor and the Ricci scalar. Since in this work we consider vacuum spacetimes $\text{Ric}\sim\mathcal{O}(\epsilon)$, the contribution of the Ricci tensor to the RHS is $\mathcal{O}(\epsilon^2)$ and we can ignore it at the order that we are considering. We are left with the following EOMs at $\mat{O}(\epsilon)$:
\begin{equation}
G_{\mu\nu} = \epsilon \left(4\,\mathcal{C}\,W_{\mu}^{\phantom{\mu}\alpha\beta\gamma}\,W_{\nu\alpha\beta\gamma}   -\tfrac{g_{\mu\nu}}{2}\,\mathcal{C}^2+8\,W_{\mu\phantom{\alpha}\nu}^{\phantom{\mu}\alpha\phantom{\nu}\beta}\nabla_\alpha\nabla_\beta\mathcal{C}\right)
\label{eq:ricci}\,,
\end{equation}
where $W_{\alpha\beta\gamma\delta}$ is the Weyl tensor since $R_{\mu\nu\rho\sigma}=W_{\mu\nu\rho\sigma} + \mathcal{O}(\epsilon)$. Then, $\mathcal{C}=W_{\alpha\beta\gamma\delta}\,W^{\alpha\beta\gamma\delta}$. System (\ref{eq:ricci}), containing derivatives up to fourth order of the spacetime metric (in $\nabla_\alpha\nabla_\beta\mathcal{C}$), has no proper classification within PDE theory. 

\begin{figure}[t!]
\includegraphics[width=0.5\textwidth]{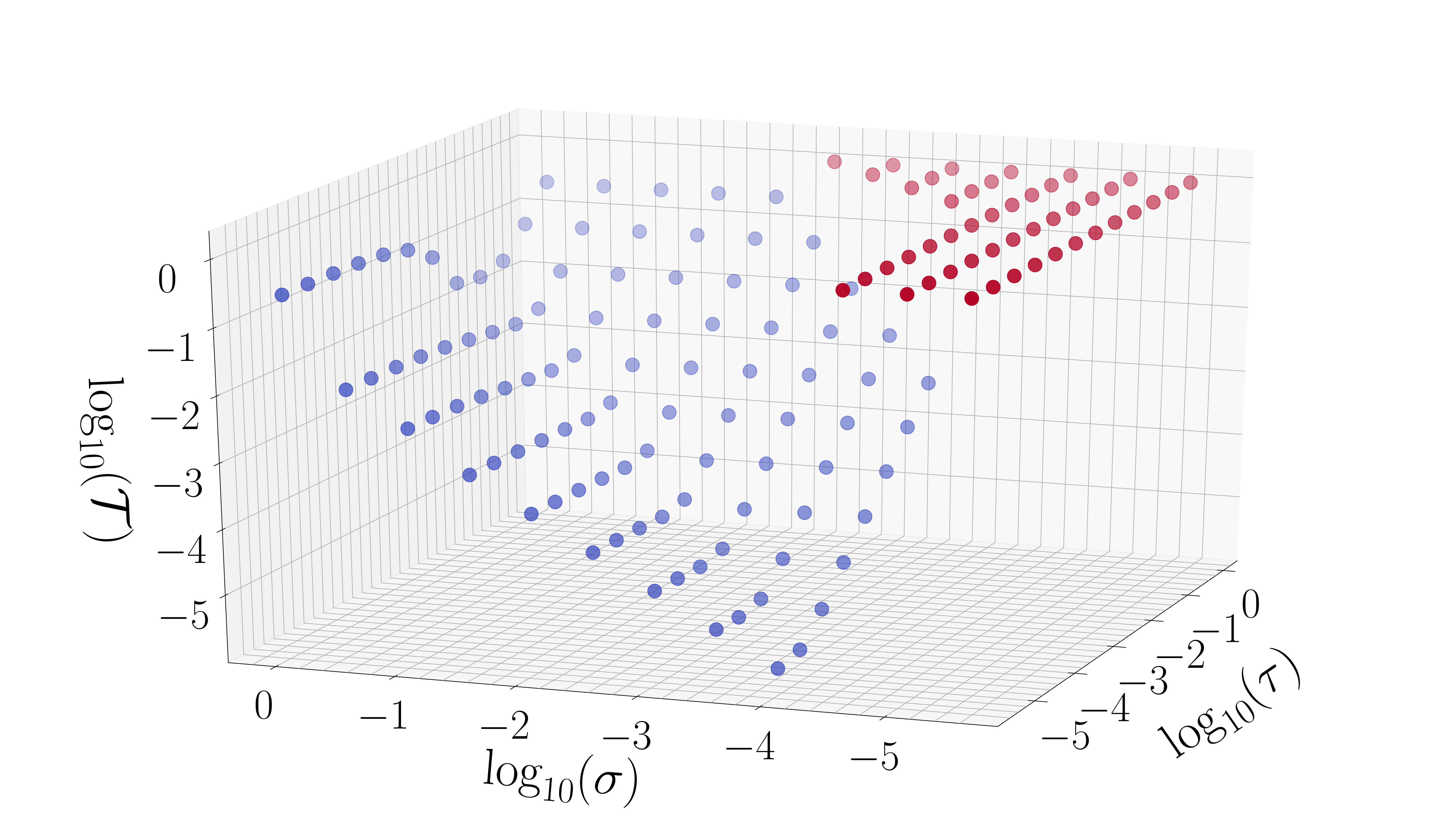}
\caption{
Tracking $\mathcal{T}$ for several values of $\sigma$ and $\tau$ for a fixed value of $\epsilon=10^{-3}$. In red we display the simulations that failed due to instabilities, and in blue the stable ones.}
\label{fig:toy_model_tracking}
\end{figure}

\PRLsection{Prototypical model}
For interacting binaries one must deal with strong curvature regions which move 
and, crucially, merge. A successful general strategy must account for the backreaction
of corrections onto the motion itself, otherwise at least secular terms will
spoil the accuracy (and hence usefulness) of the
solution~\cite{Allwright:2018rut,Okounkova:2019zjf,Reall:2021ebq}. As a preliminary
challenge, consider the following model that captures key aspects of the problem,
\begin{equation}
 \Box \phi = {-}\epsilon\, \partial^4_t \phi   \,,
\end{equation}
with $\Box$ denoting the standard d'Alembertian in Cartesian coordinates and a RHS which spoils its mathematical character. At an intuitive level one would regard the RHS as introducing small modulations
on a solution that travels at the light speed (assuming both a small parameter $\epsilon$ and regarding $\phi$ described by long-wavelength modes). However, a straightforward analysis indicates that
the higher time derivatives lead to ghost modes that grow without bound.
One can try to address this issue by `order reduction'. That is, replacing: (A) $\partial^4_t \phi \approx \partial_t^2 C(\phi)$, with $C(\phi)=\partial_{xx} \phi$ (the RHS of $\partial_t^2 \phi$ at zeroth order), or even (B) $\partial^4_t \phi \approx \partial_x^4 \phi$. Depending on the sign of $\epsilon$, Option (A) leads
to high frequency modes not propagating or blowing up, while (B) leads to even faster blowing up modes or acausal propagation. Neither is consistent with the intuition above. Generically, high frequency modes are problematic.\footnote{Recall, numerical implementations continuously feed high frequency modes through the discretization employed.} 
While in linear problems a frequency cut-off could be introduced, in non-linear ones  such a strategy is uncertain due to mode couplings potentially feeding long wavelengths into short ones, and vice versa. 
The challenge is to control the equation, render the problem of interest well posed 
and achieve a method that incorporates the effect of corrections contemplated by the theory at long wavelengths  while sensibly controlling short wavelengths; and, doing so without unduly increasing the cost of obtaining trustable solutions. 
In particular, it should allow inquiring whether a significant flow to the UV takes place which would indicate that the original theory generically abandons the EFT regime for problems of interest --unless
such UV flow takes place hidden behind a stable horizon. 
If the opposite is the case, to consistently incorporate
the effect of corrections to the theory. To address this challenge, starting from option
(A) above,\footnote{Which is the most closely associated to our desired problem.} 
we fix the equation as
\begin{eqnarray}
&&\Box \phi = {-}\epsilon\, \partial^2_t \hat C \,, \label{eq:toy_phi}\\
&&\tau  \partial_0 \hat C +
\sigma (\partial_t^2 - 2\beta^i\partial_{ti}+\beta^i\beta^j\partial_{ij})\hat{C} = C(\phi) -\hat C \,,\label{eq:toy_hatC}
\end{eqnarray}
with $\partial_0=\partial_t - \beta^i \partial_i$ and $\beta^i$ an ``advection'' vector. 
The resulting second order system determines the evolution of the 
variable $\hat C$ that is damped towards the ``source'' $C(\phi)$ on a 
timescale $\sigma/\tau$. Notice that a non-trivial stationary solution such that $C(\phi)=\hat{C}$ cannot be achieved  for non-zero values of $\tau$ and $\sigma$ (as the RHS is damped to zero, the  left hand side would continue to source it otherwise). The difference between $\hat{C}$ and its target value $C(\phi)$ decreases with $\tau$ and $\sigma$ and ultimately these parameters should be chosen to minimise this difference while preserving numerical stability.
To demonstrate the effectiveness of $\hat{C}$ accounting for $C(\phi)$ 
(which we call ``\textit{Tracking}'') and the numerical stability of the fixed system, 
we carried out a parameter exploration of $\tau$ and $\sigma$. We implement a 1D simulation in a periodic domain
of size $L=200$, discretized by a uniform grid, sixth order accurate spatial derivatives, Runge-Kutta of fourth order for time stepping, and Kreiss-Oliger dissipation with $\Delta t/\Delta x=1/4$. 
As initial data, we adopt  
$\phi(0,x)=10^{-3} \, e^{-\frac{1}{2}\left(x -100\right)^{2}}$, with $\hat{C}=\phi_{,xx}$ and $\{\phi_{,t}=\phi_{,x} ; {\hat C}_{,t}=\phi_{xxx}\}$ and fix $\epsilon = 10^{-3}$. We also
choose $\beta^i=\delta_x^i$, coincident with the speed of propagation of the main pulse
in the uncorrected equation.
During the evolution, beyond the advection
of the main `pulse', the solution develops a distinctive long oscillatory tail 
as a result of the ``correcting term'', without high frequency modes spoiling it.    
We evolved the system until the spatial extent of the tail becomes comparable to the size of the domain and compute a tracking measure as:
     $\mathcal{T}(C,\hat{C}) = \frac{||C(\phi) -\hat{C}||_{2}}{||C(\phi)||_{2}}$.
This is shown in figure \ref{fig:toy_model_tracking} for a range of $\{\sigma,\tau\}$, overall obtaining good tracking.
Notice the existence of a region where evolutions fail for values of $\sigma < 10^{-4}$ as the equations become stiff but we checked that a smaller timestep resolves this issue. Second, $\mathcal{T}$ improves linearly with decreasing $\sigma$. Third, there is a linear dependence of $\mathcal{T}$ with $\tau$ for large values, while it flattens for smaller ones and $\mathcal{T}$ depends only on $\sigma$.

\PRLsection{Gravity and Black holes}
We turn now to the demanding task of simulating dynamical BH spacetimes 
(with single boosted BHs and binaries) in the chosen EFT extension to GR.
Motivated by the previous discussion we 
{\em fix} the system \eqref{eq:ricci} by introducing a
new independent variable $\hat{\mat{C}}$ in the following way, 
\begin{align}
&G_{\mu\nu} = \epsilon \Big(4\,\hat{\mat{C}}\,W_{\mu}^{\phantom{\mu}\alpha\beta\gamma}\,W_{\nu\alpha\beta\gamma} -\tfrac{g_{\mu\nu}}{2}\,\hat{\mat{C}}^2 +8\,W_{\mu\phantom{\alpha}\nu}^{\phantom{\mu}\alpha\phantom{\nu}\beta}\,\nabla_\alpha\nabla_\beta\hat{\mat{C}}\Big)\, \label{eq:csystem_eqC} \\
&(\partial_t^2-2\beta^i\partial_{ti}+\beta^i\beta^j\partial_{ij}) \hat{\mat{C}} = \tfrac{1}{\s}\left( {\mat C}-\hat{\mat{C}} -\tau \partial_0 {\hat{\mat{C}}}\right) , \, \label{Ctt}
\end{align}
 where second time derivatives of the metric in the RHS are
replaced --following a reduction of order strategy-- using the zeroth order Einstein equations.
The resulting system of equations involves at most second-time derivatives 
and $\hat{\mat{C}}$ is damped to the physical $\mathcal{C}$ on a timescale $\simeq \sigma/\tau$ (for our choices, up to $\simeq 10 M_S$, which is shorter than any dynamical
timescales in the system). 
As a result, beyond a lengthscale $\simeq \sigma^{1/2}$ (for our choices, up to $\simeq 0.2 M_S$) the
system reduces to the original one, while shorter ones are damped and controlled, and
we can explore even shorter scales through suitable extrapolation of our results, as
we discuss in the Supplemental Material.
The new variable $\hat{\mat{C}}$, in essence, brings back (some of) the degree(s) of
freedom integrated out to arrive to the EFT theory, in this case a massive scalar (see e.g.,~\cite{Gerhardinger:2022bcw}) and our strategy defines a completion of the EFT.
Notice that the operator $\partial_0$ with the advection vector (corresponding
to the shift vector in the 3+1 decomposition) helps ensuring inflow towards
the BH(s). As we shall see, this is enough to control the whole system. That a single scalar suffices is related to it encoding the only contribution of higher derivatives and controlling it results in an overall effect ensuring high frequency modes are
kept at bay. Depending on the structure in other theories, one might need to introduce further quantities (see e.g.,~\cite{Franchini:2022ukz}). Nevertheless, the overall strategy remains unchanged.\\

\PRLsection{Initial data}
We define initial data by a single (for the single BH case) or a superposition of 
boosted BHs as described in GR and dynamically 
``turn on'' the coupling $\epsilon$ bringing it
from $0$ to the desired value with a quadratic function in a window  $t\in[10,30]M$.
This allows the coordinate conditions to settle before incorporating deviations from GR, inducing only smooth constraint violations (which are damped through the now
standard use of constraint damping \cite{Brodbeck:1998az,Gundlach:2005eh}) and by-passing the solution of initial data problem 
within the EFT theory, a task which in itself has received also limited formal and numerical attention. Again the presence of higher derivatives obscures the treatment.\footnote{See \cite{Cayuso:2020lca} for the case of a perturbed BH in spherical symmetry and \cite{Kovacs:2021lgk} for the construction of initial data in scalar-tensor theories of gravity.}

For initial data in the single BH case,
we use a boosted BH solution derived from the conformal transverse-traceless
decomposition \cite{Lichnerowicz:1944,York:1971hw,York:1972sj}, which uses an approximate conformal
factor solution to the Hamiltonian constraint, valid for small boosts.
For the binary BHs, we adopt Bowen-York-type-of initial data \cite{Bowen:1980yu} describing two superposed
equal mass, boosted, non-spinning BHs in a quasi-circular orbit.  The individual masses are $m_i\approx 0.5M$, $i=1,2$, and the separation is $D\sim 12M$ (initial orbital
frequency $\simeq 0.025/M$). The momenta are tuned so this binary is initially
in quasi-circular motion, and in GR it describes 12 orbits before merger (the initial
BHs velocities are similar to those in the single boosted BH case).

\PRLsection{Evolution}
We use the \texttt{GRChombo} code \cite{Clough:2015sqa,Andrade:2021rbd} and the CCZ4 formulation of the Einstein equations \cite{Alic:2011gg} (see also \cite{Bernuzzi:2009ex}) which implements the system \eqref{eq:csystem_eqC}-\eqref{Ctt} with a distributed
adaptive mesh refinement capabilities,\footnote{Here we use a 2:1 mesh refinement ratio.} using $6^\text{th}$ order finite difference operators for
the spatial derivatives and the method of lines for time integration through a Runge Kutta of $4^\text{th}$ order.\footnote{We use $6^\text{th}$ Kreiss-Oliger dissipation with a dissipation coefficient  $\sigma_\text{diss}=2$ (e.g.,~\cite{Calabrese:2003vx}).} In this way, we only need to use three ghost cells along each coordinate direction.
We adopt
a standard 1+log slicing condition for the lapse $\alpha$ and the Gamma-driver for the shift $\beta^i$, as implemented in the public version of the code and adopt
Sommerfeld boundary conditions at the outer boundaries.
We redefine the damping parameter $\kappa_1\to\kappa_1/\alpha$ to ensure that it remains active inside the apparent horizon (AH) \cite{Alic:2013xsa}; we decrease the damping parameter in the shift condition as well as increase $\sigma$ in \eqref{Ctt} at large distances from the centre of the binary to ensure that no violation of the CFL condition arises due to grids becoming coarser  with our explicit time-stepping strategy (e.g.,~\cite{Schnetter:2010cz}).
Otherwise, the chosen values for the constraint damping, shift and lapse conditions are 
$\{\kappa_1=1,\kappa_2=-0.8,\kappa_3=1,\alpha_2=\alpha_3=1,\alpha_1=2,\eta_1=0.75,\eta_2=1/M\}$ (\cite{Clough:2015sqa}). In the code, the additional evolution equation \eqref{Ctt} is implemented in the obvious first order form. 
We excise a region of the interior of BHs as in \cite{Figueras:2020dzx} which removes the role of correcting terms, achieving stable evolutions
with unduly high resolution (see the Supplemental Material for more details).

With the total ADM mass $M$ of the system setting a scale, our
domain for the boosted BH case corresponds to the quadrant $x\in [-L,L]$ and $y,z\in[0,L]$ as symmetry allows for restricting it.
We adopt $L=384 M$ with a the coarsest grid spacing (for production runs) $\Delta=2 M$; we then add another 6 levels of refinement. For the BH binary case the computational domain, exploiting symmetries, is given by $x,y\in[-L,L]$ and $z\in[0,L]$.  In this case we adopt, $L=512M$ and coarsest grid spacing (for production runs) $\Delta=4 M$; we then add another 8 levels of refinement (for convergence
tests we consider up to $\Delta=8/3 M$ and same number of refined levels). We extract the gravitational waves at 6 equally spaced radii between $R=50M$ and $R=100M$ and extrapolate the result to null infinity. 
In both cases we use the second (spatial) derivatives of the conformal factor\footnote{Recall that the conformal factor $\chi$ is one of the evolution variables in the CCZ4 formulation and it is related to the induced metric on the spatial slices $\gamma_{ij}$ as $\chi=1/(\det\gamma)^\frac{1}{3}$.} $\chi$ to estimate the local numerical error and determine whether a new level of refinement needs to be added; in addition, we fix the spatial extent of certain refinement levels to ensure that the resolution at the chosen extraction radii is high enough. 
Lastly, we choose a scale value $|\epsilon|=10^{-5}M_S^{6}$, which implies
a coupling scale for new physics beyond GR of $\Lambda\approx 7/M_S \simeq 5 M_{\odot}/M_S$ km$^{-1}$. Notice for this small scale, correction effects will be undoubtedly subtle,
with a consequent high accuracy requirements to capture them.  Here  we undertake a first
study mainly focused on demonstrating the ability of the method to control the system. We
will concentrate on assessing this and obtaining a qualitative description of observed
consequences.
We consider both signs for $\epsilon$, the negative case satisfies the
constraints argued for in~\cite{deRham:2021bll}, the positive one also provided
azimuthal numbers of the solution are not large, which is our case. We here choose a conservative scale $M_S= 10 M_{\odot}$, i.e., somewhat below (but comparable)
to the curvature scale set by the masses of the individual BHs for all detected gravitational wave events.  
Choosing a smaller scale would imply that the modifications become ${\cal O}(1)$ during the inspiral~\cite{Sennett:2019bpc} with arguably clear imprints on
the observed signal, which is  inconsistent with observations.  
Further, we note that it is
natural to expect the scale to remain fixed, thus the larger  the BH
mass, the smaller the effect of corrections would be. This observation is particularly relevant
as the BHs merge, as corrections after such regime would naturally become smaller.\footnote{Data analysis techniques can
exploit these observations (e.g.,~\cite{Silva:2022srr,Dideron:2022tap}).} Since for larger masses corrections would be smaller, we here
focus on masses comparable to the length scale $M_S$ 
we thus adopt individual BH masses $m_i=M_S/2=5 M_{\odot}$.

\begin{figure}[t!]
\includegraphics[width=0.475\textwidth]{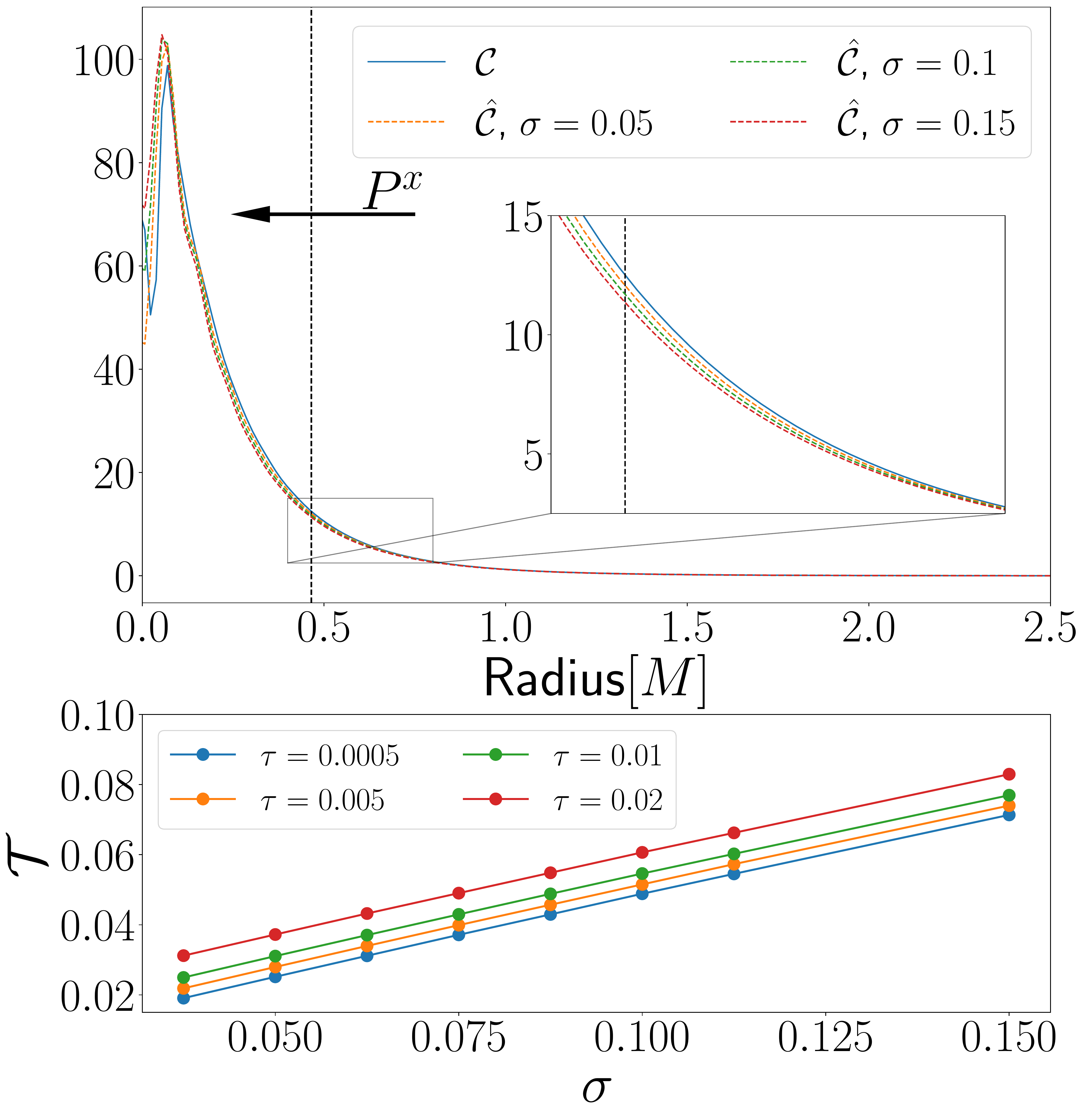}
\caption{Top: $\mathcal{C}$ and $\hat{\mathcal{C}}$ on a line starting from the puncture and trailing the boosted BH --where differences in tracking are larger-- (at $t=75M$, when all transients related to gauge and initial data have passed) for a fixed value of $\tau =0.005$. The BH has mass $M=0.5$ and momentum $P^{x} = 0.08M$. The vertical dashed black line denotes the location of the AH and the arrow indicates the direction of motion of the BH. Bottom: $\mathcal{T}$ for a range of $\{\sigma,\tau\}$ computed over the profile displayed in the top figure outside the AH up to $r=5M$ (the results do not change by a larger integration domain).}
\label{fig:single_bh_tracking}
\end{figure}

\PRLsection{Single boosted black holes}
We confirmed our strategy's ability to evolve boosted (and stationary) BHs, with
the solution reaching a steady state behavior shortly after the corrections are fully
turned on. The solution is smooth without inducing growth in high frequency modes or signs of instability. 
Beyond GR effects are naturally larger in the BH region. By comparing the value of $\hat{\mat{C}}$ and $\mathcal{C}$ we confirm the former tracks the physical one quite well
and that lower values of $\{\sigma,\tau\}$ improve the tracking behavior. 
Figure~\ref{fig:single_bh_tracking} illustrates the observed behavior of $\mathcal{T}(\mat{C},\hat{\mat{C}})$. Importantly, examination of the relative difference between two values of $\hat{\mat{C}}$  obtained with two different values of $\sigma$
(and analogously with $\mathcal{C}$) indicates errors associated to the choice of this
parameter do not severely accumulate, thus the solution is not degraded by strong
secular effects (see Fig.~\ref{fig:errorvstime}). For instance, it would 
take $\approx 10^6 M$ for 
the  relative error for Kretschmann scalar with $\sigma=0.1$ and $0.05$ to be
of order ${\cal O}(1)$. 
Numerical instabilities develop for smaller values of $\sigma$ around the excision region, well inside the AH; these instabilities are sensitive to the details of the excision --improving with resolution. This suggests other forms of excision would be more robust 
as $\sigma$ is decreased (e.g.,~\cite{Cayuso:2020lca}). With a successful handling of correction effects in single, moving BHs, we turn next focus on the
challenging setting of binary BH merger. 

\PRLsection{Black hole binary mergers}
The binary tightens due to emission of gravitational waves which radiate energy
and angular momentum from the system.
Figure~\ref{fig:binary_bh_tracking} shows the gravitational wave strains
for different values of $\epsilon$ and contrasts them with the corresponding one in GR.
The solution is smooth, without any signs of instability throughout inspiral,
merger and ringdown. The corrections to GR and their high degree of non-linearity
and higher gradients contributions certainly tax resolution requirements. Our
studies here are not focused on quantitatively sharp answers, but on testing
the approach with enough resolution for qualitatively informative results and 
contrasting them with results in GR.
In particular, we see that positive/negative values of $\epsilon$ induce a (slight) merger phase delay/advance. This is consistent with expectations, 
where BHs in this theory have non-zero tidal effects, encoded
to leading order in the Tidal Love Number $\kappa \propto \epsilon$ ~\cite{Cardoso:2018ptl}.
The binary behavior in the 
inspiral regime can be captured through a Post-Newtonian analysis which shows tidal effects induce a phase offset $\propto - \kappa$ (hence $\propto - \epsilon$) ~\cite{Flanagan:2007ix} (see also~\cite{Porto:2016zng,Endlich:2017tqa}) for BHs with size comparable to $M_S$. Leading order Post-Newtonian
estimates for the phase difference give $\approx \pm 5 \times 10^{-3}$ radians up to a
common gravitational wave frequency ($M f=0.01$) for negative/positive values of $\epsilon$ in Fig.~\ref{fig:binary_bh_tracking}. Our obtained offsets --extrapolated to $\sigma \rightarrow 0$--  are consistent with the sign, though about $200$ times larger 
(for a related study in Einstein-Scalar-Gauss
Bonnet theory see~\cite{Corman:2022xqg}).

The BHs coalesce and the resulting peak strain is 
comparatively similar to that in GR and no significant further structure is induced in the multipolar decomposition of the waveforms,  confirming that the solution stays within the EFT regime. Moreover,
since the merger gives rise to a BH with roughly twice the
individual masses, corrections are reduced by $\approx 2^{-6}$. Thus the final BH is closer to a GR
solution than the initial ones. After
the peak amplitude, the system settles quickly into a stationary BH
solution. This transition is described by an exponential, oscillatory behavior
described by quasi-normal modes (QNM). {While QNM spectra have only been computed for
slowly rotating BHs in this theory \cite{Cardoso:2018ptl,Cano:2020cao}, the departure observed in decay rates is consistent with extrapolation to higher spin values, though this is not the case in the oscillatory frequency. We note, however, that
the extracted values for the case in GR, have relative errors 
$\approx 0.1\%$; since
GR corrections to the QNMs in the case studied here 
are subleading by an order of magnitude such potential discrepancy can
be attributed to a need for even higher accuracy to capture them sharply.

Also, as the system approaches its stationary 
final state we confirm it is axisymmetric. Such symmetry is expected in stationary BH solutions in EFTs of gravity~\cite{Hollands:2022ajj}.  By evaluating
different scalars, such as the conformal factor $\chi$ of the spatial metric
and $\hat{\mat{C}}$, on the intersection of the AH (which in the stationary case coincides with the event horizon) with the equatorial plane,
the tendency towards axisymmetry can be confirmed (see figures \ref{fig:C_diff_axisymmetric},\ref{fig:chi_diff_axisymmetric}).
Last, note that the difference in the innermost stable
circular orbit frequency between slowly rotating BHs in this theory and GR
goes as  $\delta \Omega_{\text{ISCO}}\propto -\epsilon$. Thus, extrapolating
this observation to general spins,  and following 
the successful strategy to estimate the final (dimensionless) spin in BH
coalescence in GR~\cite{Buonanno:2007sv}, one can argue that the final
BH spin should be higher/(lower) for positive/(negative) values of $\epsilon$ as
the final `plunge' takes place with a higher/(lower) contribution of orbital
angular momentum to the final BH.
Cautioning that a higher accuracy is required to confirm this expectation,
our results are consistent with it. 

\begin{figure}[t!]
\includegraphics[width=0.475\textwidth]{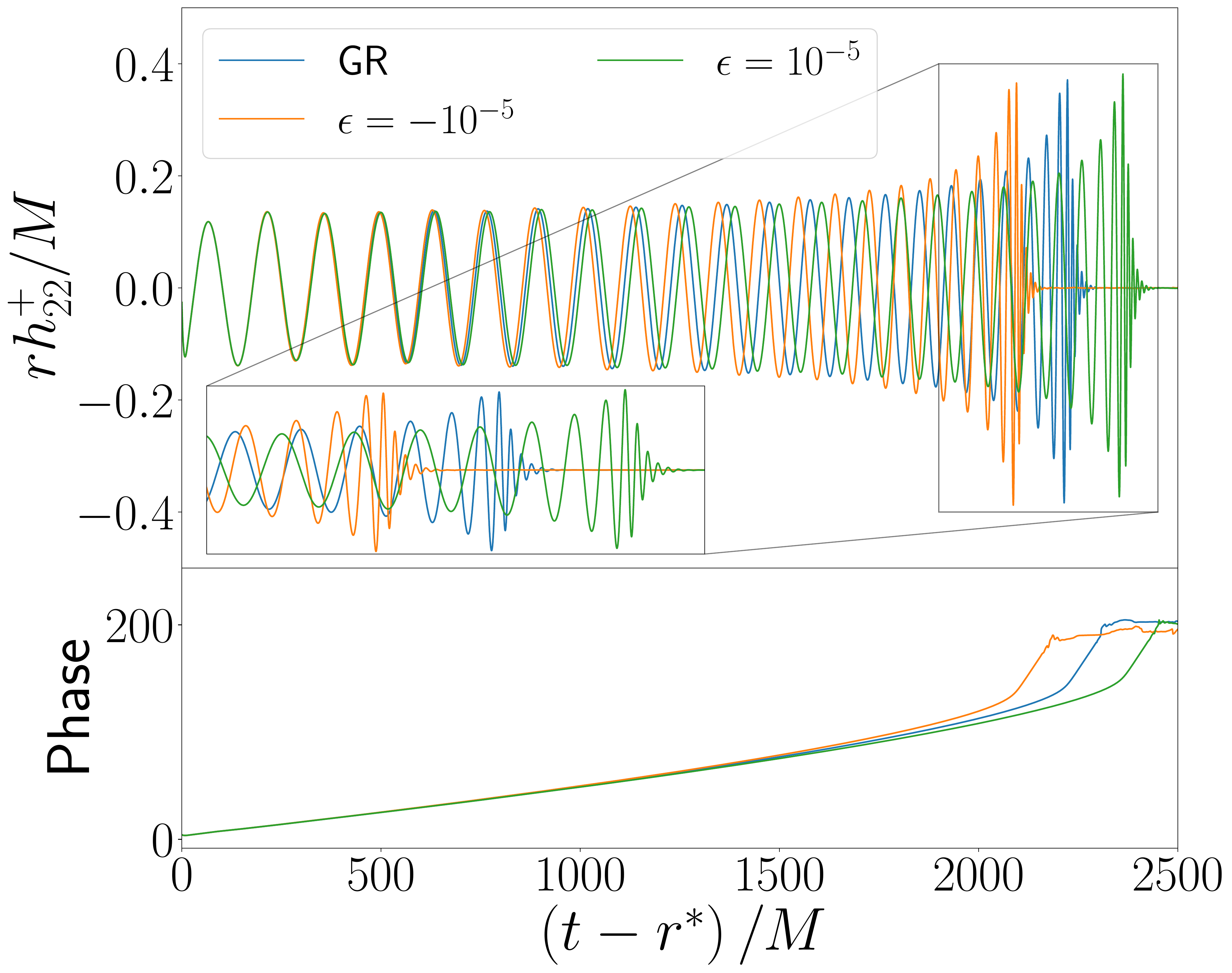}
\caption{Gravitational waves in GR and in the EFT theory ($\epsilon=\pm 10^{-5}$) 
with $\sigma=0.0625,\tau=0.005$. 
Top: $\left(\ell,m\right)=\left(2,2\right)$ mode of the $+$ polarization $h^{+}_{22}$, extrapolated to null infinity as a function of the retarded time $u = t -r^{*}$, where $r^{*}$ is the tortoise radius. Bottom: gravitational wave phase of $h_{22}$.} 
\label{fig:binary_bh_tracking}
\end{figure}

\PRLsection{Discussion}
We have demonstrated the ability of the ``fixing'' approach to
enable studies of beyond GR theories. This approach, in
particular, provides
a practical way to explore phenomenology in the highly non-linear and dynamical regime of compact binary mergers. Especially relevant is that it enables assessing whether the solution for cases of interest remains in the EFT regime and
the impact of corrections in gravitational wave observations. From this first set of analysis, we conclude the solution does remain in this regime for comparable mass, quasicircular mergers in the 
observable region, i.e., the BH(s) exterior. Thus, much like in the case
of GR, a strong UV energy flow takes place inside the horizon
but not in the outside region, staying
within the valid EFT regime.
While in the current work we have focused on a specific theory and scale,
our choice was motivated by stress-testing the approach with
highly demanding challenges --brought by higher than second order derivatives
and in the context of BH collisions. However, the underlying strategy is applicable
also in beyond GR theories with second order equations that
can induce change of character in the equation of motion (e.g.,~\cite{Ripley:2019hxt,Franchini:2022ukz}). We note that for a particular class of non-linear theories (with second order equations
of motion) consistent, non-linear studies have been presented~\cite{Corman:2022xqg,AresteSalo:2022hua,Bezares:2021dma}. However, they
required significant supporting theoretical efforts to identify appropriate gauge
conditions and merging BH solutions have been obtained up to 
some maximum coupling value otherwise mathematical pathologies arise.
Our approach, in principle, provides a way to robustly explore beyond such
coupling and, in general, study beyond GR theories self-consistently where such
supporting theoretical input is not available or even achievable without strong --and a priori unjustifiable-- assumptions. 
Of course, practical application of the approach described here
should be mindful of checking results upon variations of
ad-hoc parameters to ensure, given a coupling length, scales
are sufficiently resolved.

\PRLsection{Acknowledgements}
We thank William East, Rafael Porto, Oscar Reula
and Robert Wald for discussions.
This work was supported in part by
Perimeter Institute for Theoretical Physics. Research
at Perimeter Institute is supported by the Government
of Canada through the Department of Innovation, Science and Economic Development Canada and by the
Province of Ontario through the Ministry of Economic
Development, Job Creation and Trade. LL thanks
financial support via the Carlo Fidani Rainer Weiss Chair
at Perimeter Institute. LL receives additional financial support from the Natural Sciences and Engineering
Research Council of Canada through a Discovery Grant and CIFAR. PF is supported by Royal Society University Research Fellowship Grants  RGF\textbackslash EA\textbackslash 180260, URF\textbackslash R\textbackslash 201026 and RF\textbackslash ERE\textbackslash 210291. TF was supported by a PhD studentship from the Royal Society RS\textbackslash PhD\textbackslash 181177.  The simulations presented used PRACE resources under Grant No. 2020235545, PRACE DECI-17 resources under Grant No. 17DECI0017, the CSD3 cluster in Cambridge under projects DP128 and DP214, the Cosma cluster in Durham under project DP174 and the ARCHER2 UK National Supercomputing Service under project E775. The Cambridge Service for Data Driven Discovery (CSD3), partially operated by the University of Cambridge Research Computing on behalf of the STFC DiRAC HPC Facility. The DiRAC component of CSD3 is funded by BEIS capital via STFC capital Grants No. ST/P002307/1 and No. ST/ R002452/1 and STFC operations Grant No. ST/R00689X/1. DiRAC is part of the National e-Infrastructure.\footnote{\texttt{www.dirac.ac.uk}} The authors gratefully acknowledge the Gauss Centre for Supercomputing e.V.\footnote{\texttt{www.gauss-centre.eu}} for providing computing time on the GCS Supercomputer SuperMUC-NG at Leibniz Supercomputing Centre.\footnote{\texttt{www.lrz.de}} This research was also enabled in part 
by support provided by SciNet (www.scinethpc.ca) and Digital Research Alliance of Canada (alliancecan.ca).

\appendix
\section{Supplemental Material}

\PRLsection{Excision}
The numerical treatment of solutions containing BHs
requires dealing with the singular behavior of its interior.
In GR, excising from the computational domains  trapped region(s) 
(which can be shown to lie within BHs and thus are causally disconnected from its exterior) or mapping
the interior of BHs to other causally disconnected asymptotic regions are practical and 
successful techniques to address this issue. 
With beyond GR theories, extensions of
these ideas can be adopted.
Here, we follow a strategy implemented in \cite{Figueras:2020dzx}, where terms
beyond GR are ``turned off'' inside the AH, thus
allowing one to follow the standard approach in GR and use puncture gauge to handle (coordinate) singularities inside BHs.  By turning off beyond GR terms well inside the AH we are modifying the theory in a region of spacetime that is causally disconnected from any external observer and where the theory itself should no longer be a valid EFT anyway. 

Since the higher-derivative terms enter the equations of motion as an effective stress-energy tensor $T_{\m\n}$, a measure of the weak field condition is provided by the ``size'' of the components of $T_{\m\n}$.  In the 3+1 decomposition, we have 
\begin{equation}
    \r=n^\m n^\n T_{\mu\nu}\,,\quad S_i=-n^\m\gamma_{i}^{\phantom{i}\n}T_{\m\nu}\,,\quad S_{ij}=\gamma_{i}^{\phantom{i}\m}\gamma_{j}^{\phantom{i}\n}T_{\m\n}\,.
\end{equation}
Then, a point-wise measure of the weak field condition is captured by the quantity
\begin{equation}
    W = \sqrt{\r^2 + S_i S_j \d^{ij} + S_{ij} S_{kl} \d^{ik} \d^{jl}}
    \label{eq:def_weak}
\end{equation}
This quantity need not be covariant nor anything special, it is just a measure where we can input a threshold; it is preferred to $\r$ which goes to zero at the puncture (even though it is large in the surrounding region).\footnote{This is because we turn off the non-GR terms inside BHs, as we explain in the following paragraph.}

Given an energy-momentum tensor, we damp the ``effective source'' inside the AH and when  the weak field condition $W$ is too large  by a factor $e\cdot T_{\m\n}$  where $e$ is a smooth transition function:
\begin{align}
    e(\chi,W) &= 1 - \s(\chi; \bar{\chi}, \w_\chi) \s(W; \bar{W}, -\w_W)\,,\\
    \s(x; \bar{x}, \w_x) &= \frac{1}{1 + 10^{\frac{1}{\w_x}\br{\frac{x}{\bar{x}}-1}}}\,,
\end{align}
where  $\chi$ is the conformal factor of the induced metric, $\bar{\chi}$ and $\bar{W}$ are thresholds for the excision cutoff region and $\w_\chi$ and $\w_W$ are smoothness widths. With this choice of $e(\chi,W)$, the source approaches 0 exponentially whenever $\chi < \bar{\chi}$ and $W > \bar{W}$,  and is 1 otherwise. The parameters $\w_\chi$ and $\w_W$ determine the width of the transition region; we typically choose them to be $\w_\chi=\w_W=0.1$. Since in our working gauge the contours of $\chi$ track the AH very well (see the discussion below), we choose a threshold $\bar{\chi}$ that ensures that the excision region is well inside the AH. In practice we ended up taking $\bar{W} = 0$ so that $\s(W; \bar{W}, -\w_W)$ is always effectively $1$ and hence the excision region is solely determined by the value of $\chi$. In practice, for non-spinning BH binaries,  the choice $\bar{\chi} = 0.09$ ensures that the excision region is well within the AH \cite{Radia:2021smk, tiago_thesis}.

\PRLsection{Impact of ad-hoc parameters}
As described in the main text, the parameters $\{\sigma,\tau\}$ appearing in the evolution equation for $\hat{\mat{C}}$ \eqref{Ctt}
are ad-hoc and define a scale.\footnote{The analogous parameters appearing in Israel-Stewart formulations of relativistic viscous hydrodynamics can be related to transport coefficients of the underlying microscopic theory.} Their role is to control scales
shorter (heavier) than theirs thus controlling the mathematical pathologies 
discussed. Naturally, the solution obtained for any set of values of 
such parameters can  depend on their values. Such dependency will
be strong if the solution displays a significant cascade to the UV, as in this
case the system tends to abandon the EFT regime and the fixing strategy would
force it to remain in it, damping energy in high frequency modes.
On the other hand, such dependency could be mild, not affecting qualitatively
the solution's behavior but might introduce minor variations in quantitative
characteristics. Whenever the fixing strategy is employed (and arguably any other
strategy) to address the mathematical shortcomings, 
assessing the solution's dependency on whichever strategy is paramount. In
our case, it means examining dependency on $\{\sigma,\tau\}$.

By varying their values, we do identify a mild depdendecy on them,
which we trace back to the role of the advection vector $\beta^{i}$ and the terms it
multiplies in \eqref{Ctt}, and not to a UV cascade taking
place outside the BH(s). These terms, which were
introduced to ensure advection into the BH, also affect to a small
degree the tracking, as we have seen. In particular, the difference  between $\hat{\mathcal{C}}$ and $\mathcal{C}$ is linear, and increasing with $\sigma$ 
especially where curvature is strong. This difference, although small, over long times affects somewhat physical quantities or observables. One can nevertheless, extrapolate
such difference to values of $\tau,\sigma \rightarrow 0$ 
(see also \cite{Bezares:2021yek}). For instance, as displayed in figure \ref{fig:extrapolated_peak_amplitude} we perform such extrapolation to obtain the time of peak amplitude of the strain for $\sigma\rightarrow 0$. In this figure we show
such quantity depends linearly with $\sigma$ and the extrapolated time delay for the peak amplitude is $\approx 30M$. This extrapolation is further supported by the weak dependence on $\tau$ and the fact that a linear behaviour with $\sigma$ is present in our toy model until very small values of this parameter. 
Further, we can monitor the difference of between values of $\hat{\mathcal{C}}$, or $\mathcal{C}$, upon varying $\sigma$ as
time proceeds. Figure~\ref{fig:errorvstime} illustrates the 
observed behavior for the single boosted BH case, illustrating that errors related to the value of
$\sigma$ accumulate slowly. Indeed, it would take
about $t\approx 10^6 M$ for the relative error
for Kretschmann scalar with $\sigma=0.1$ and $0.05$
to be ${\cal O}(1)$.

\begin{figure}[t!]
\includegraphics[width=0.45\textwidth]{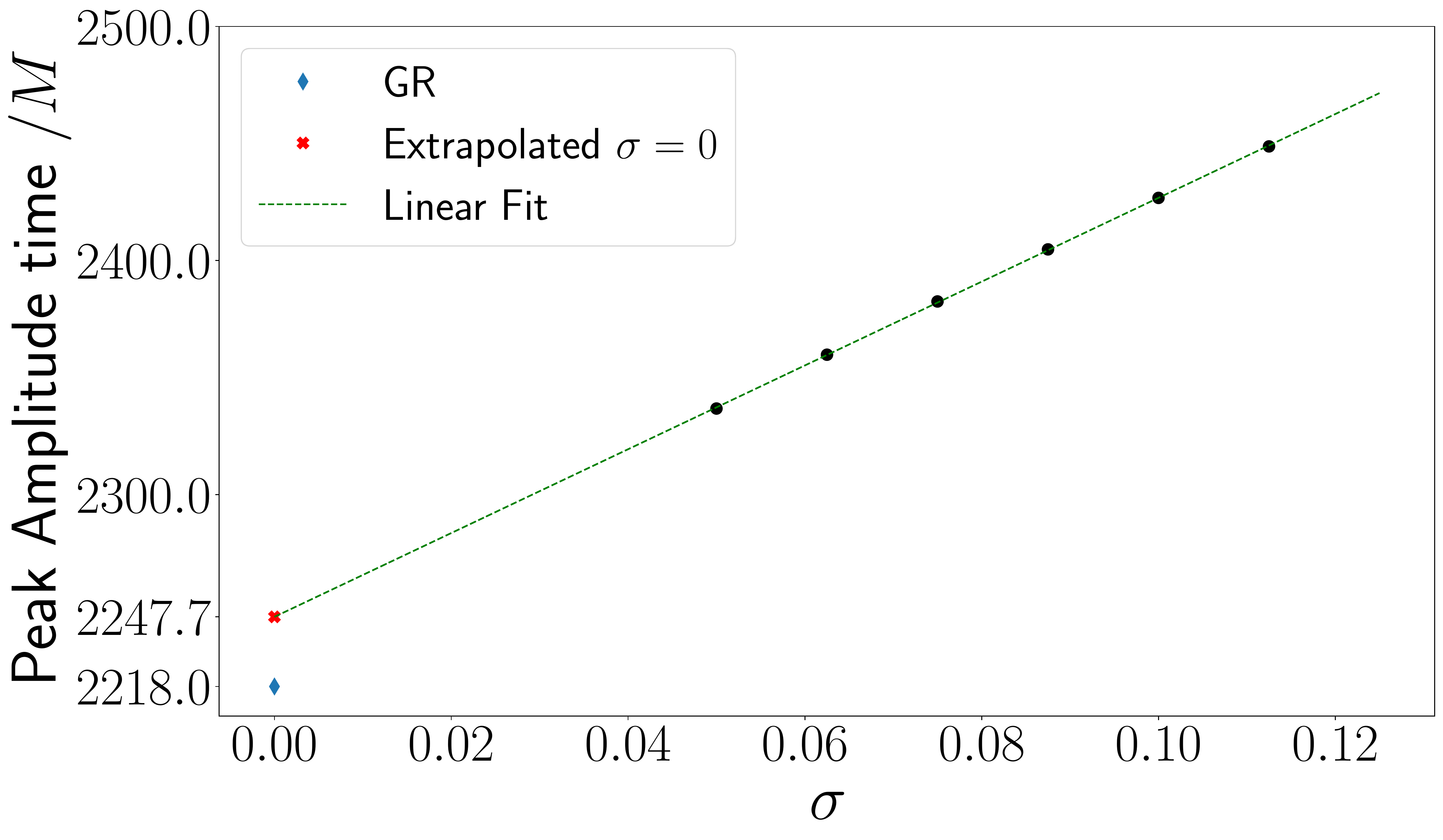}
\caption{Retarded time for peak amplitude of $h_{22}$ of BH binary mergers for a fixed value of $\tau=0.005$ and several values of $\sigma$ (black dots). The dashed green line is a linear fit to the data given by the black dots. The red cross is a linear extrapolation for the peak amplitude time for $\sigma=0$ which has a value of $t^{peak}_{\sigma \rightarrow 0} = 2247.7 M $, while the blue diamond is the peak amplitude time for the GR case and has a value of $t^{peak}_{GR}=2218.0M$.} 
\label{fig:extrapolated_peak_amplitude}
\end{figure}

\begin{figure}[t!]
\includegraphics[width=0.45\textwidth]{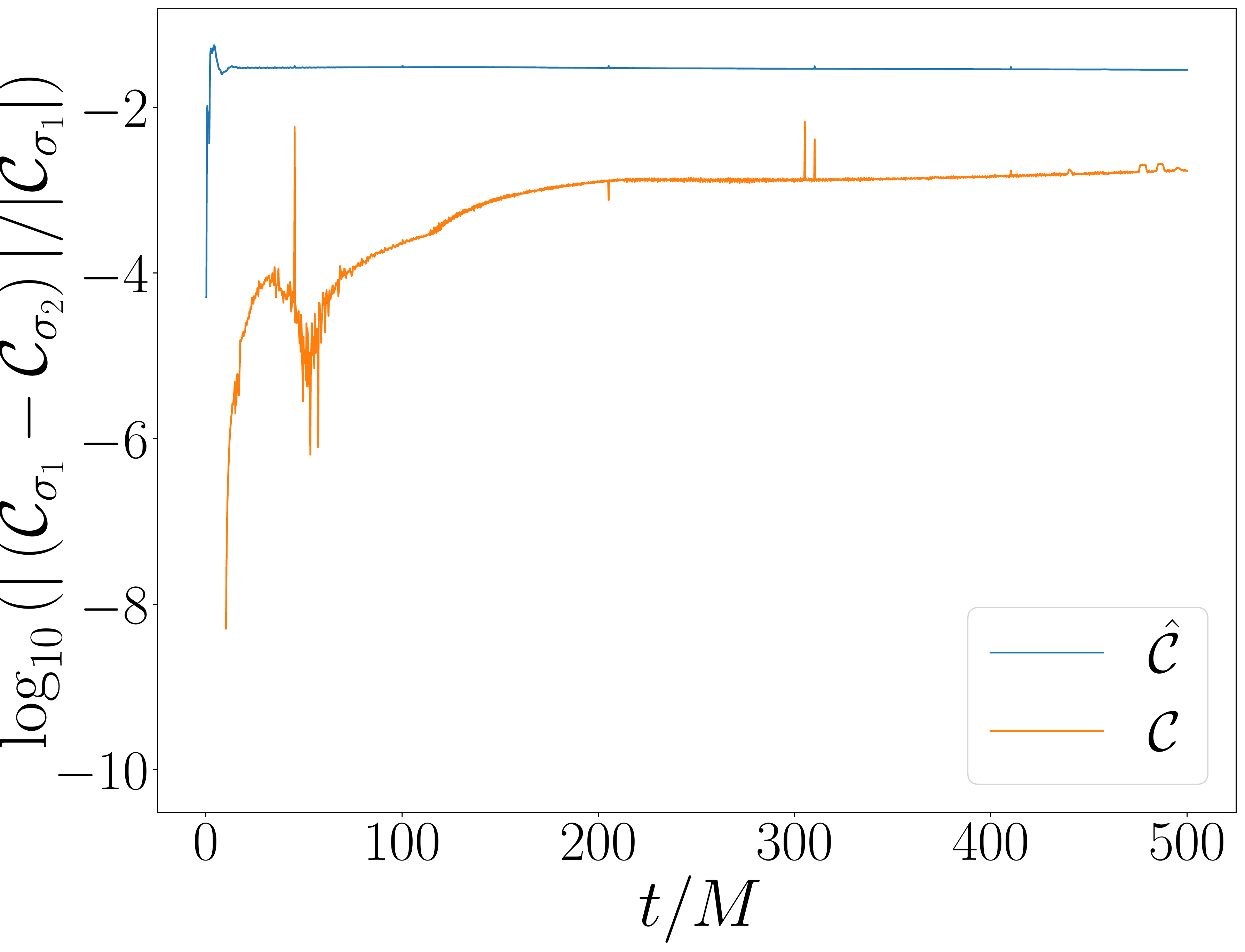}
\caption{Relative difference between the maximum value of $\hat{\mathcal{C}}$ ($\mathcal{C}$) on the AH for single boosted BHs for two values of $\sigma$ ($\sigma_{1}=0.05$ and $\sigma_{2}=0.1$).} 
\label{fig:errorvstime}
\end{figure}

\PRLsection{Axisymmetry of the remnant Black Hole}
To assess the axisymmetry of the final state, we evaluate two scalar quantities, namely the  the conformal factor $\chi$ on the $t=\text{const}$ slices
and $\hat{\mathcal{C}}$, at the intersection of the common AH with
the equatorial plane and compute their relative differences with the (arbitrary)
point $\phi=0$, where $\phi$ is the usual azimuthal angle on the AH two-sphere. As time progresses, such difference reduces significantly
indicating a high degree of axisymmetry, consistent with the result of~\cite{Hollands:2022ajj}.

\begin{figure}[t!]
\includegraphics[width=0.45\textwidth]{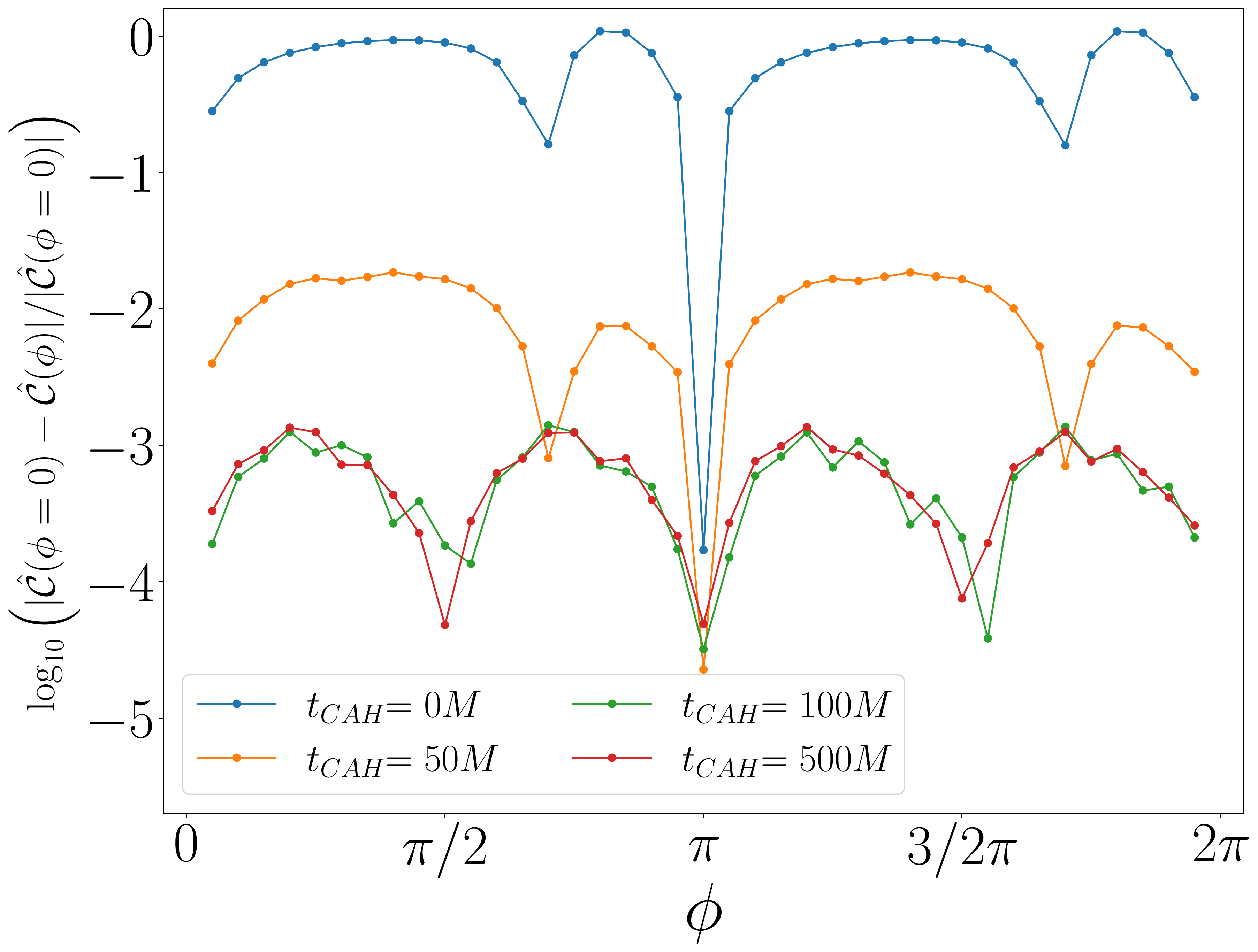}
\caption{Relative difference between $\hat{\mathcal{C}}(\phi)$ and $\hat{\mathcal{C}}(\phi=0)$ evaluated at different azimuthal angles $\phi$ at the equator of the common AH formed after merger at different times starting from such time at which a first
common AH is found ($t_{CAH}=0$). The configuration evolves towards an axisymmetric state.} 
\label{fig:C_diff_axisymmetric}
\end{figure}

\begin{figure}[t!]
\includegraphics[width=0.45\textwidth]{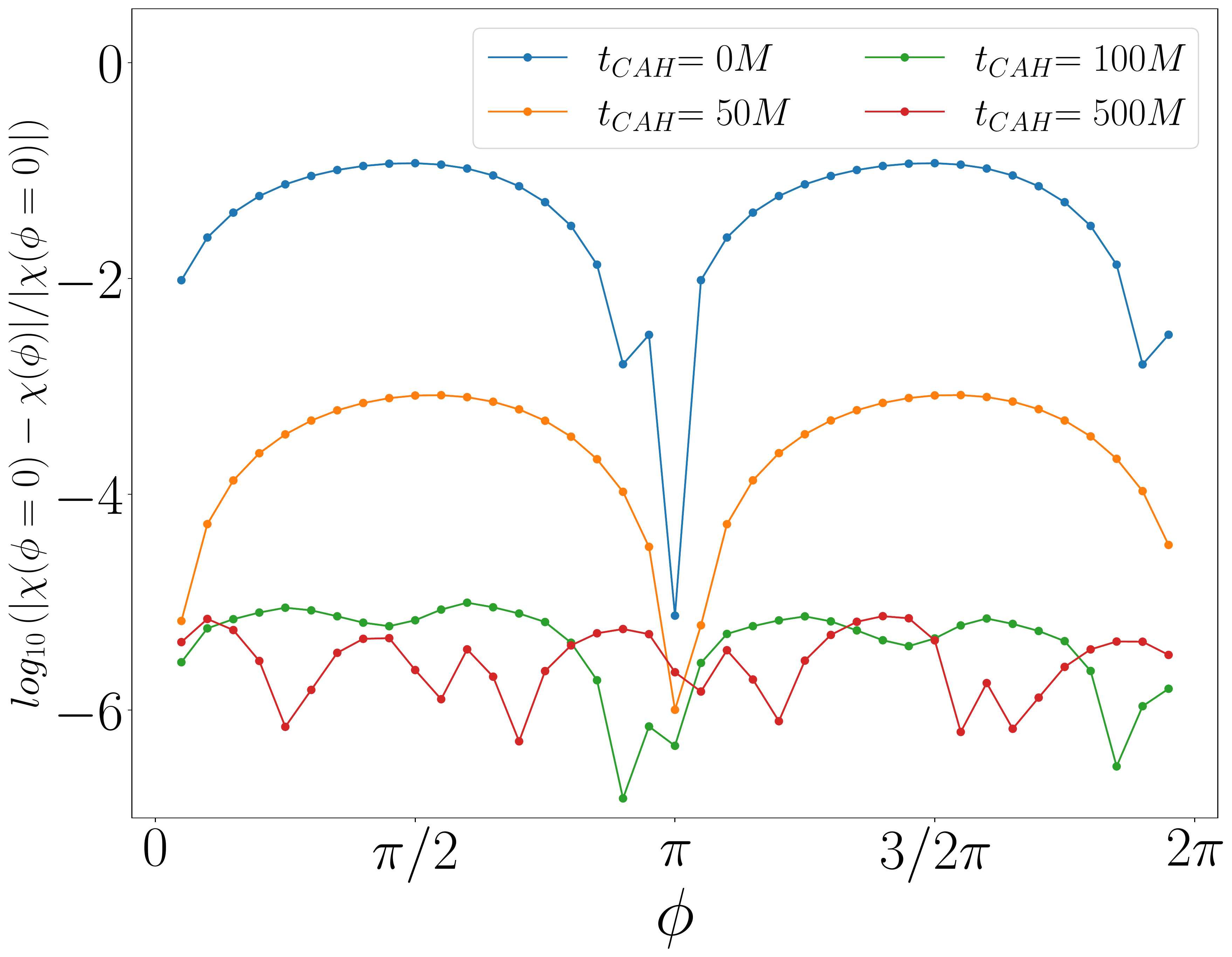}
\caption{Relative difference between the conformal factor $\chi(\phi)$ and $\chi(\phi=0)$ as in Fig.~\ref{fig:C_diff_axisymmetric}. As time progresses
a clear tend to axisymmetry is evident.} 
\label{fig:chi_diff_axisymmetric}
\end{figure}

\PRLsection{Convergence}
We test convergence by comparing the evolution presented, which used a numerical grid coarsest spacing of $\D = 4M$ and 8 further levels of refinement, with two higher resolutions, with spacing $\D = \tfrac{32}{11}M$ (medium) and $\D = \tfrac{8}{3}M$ (high). The particular setting for these simulations has $\epsilon = 10^{-5} M_{S}^6$, $\sigma=0.1$ and $\tau=0.005$.  Figure \ref{fig:gw_convergence} shows gravitational strain errors between low, medium and high resolutions, together with the estimates for second and fourth order convergence respectively based on the convergence proportionality factor
\begin{equation}
    Q_n = \frac{\D_\text{Low}^n-\D_\text{Med}^n}{\D_\text{Med}^n-\D_\text{High}^n}\,.
\end{equation}
This figure indicates approximately fourth order convergence for both the amplitude and the phase of $h^{+}_{22}$.

\begin{figure}[t!]
\includegraphics[width=0.5\textwidth]{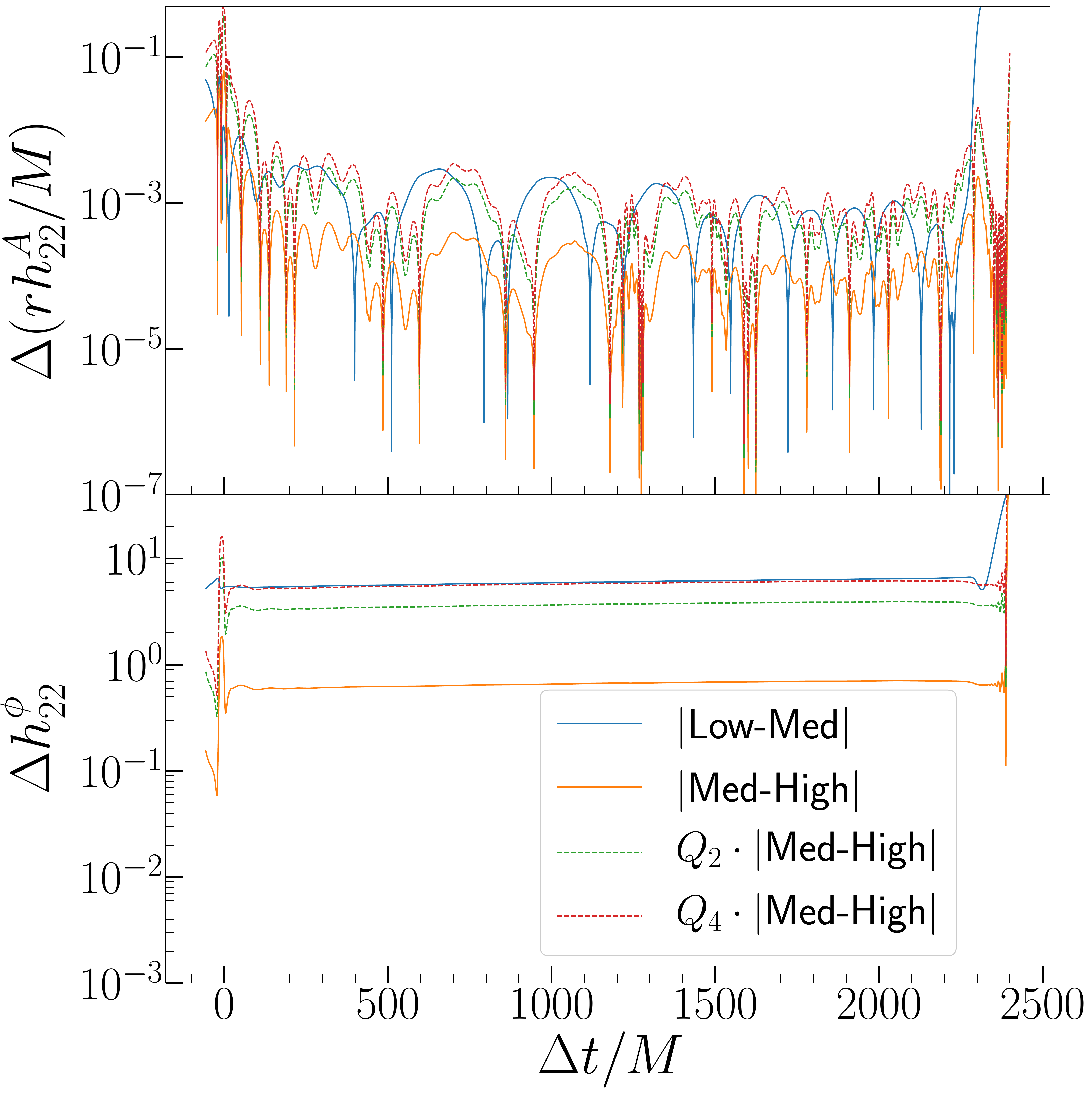}
\caption{Errors for the amplitude (top) and phase (bottom) of the $+$ polarisation $h^{+}_{22}$, extrapolated to null infinity, as a function of retarded time. The dotted lines show estimates for the error between low and medium resolutions assuming second and fourth order convergence. The waves were aligned by their peaks. These figures are consistent with approximately fourth order convergence.}
\label{fig:gw_convergence}
\end{figure}




\bibliographystyle{apsrev4-2}
\bibliography{main}

\begin{thebibliography}{52}%
\makeatletter
\providecommand \@ifxundefined [1]{%
 \@ifx{#1\undefined}
}%
\providecommand \@ifnum [1]{%
 \ifnum #1\expandafter \@firstoftwo
 \else \expandafter \@secondoftwo
 \fi
}%
\providecommand \@ifx [1]{%
 \ifx #1\expandafter \@firstoftwo
 \else \expandafter \@secondoftwo
 \fi
}%
\providecommand \natexlab [1]{#1}%
\providecommand \enquote  [1]{``#1''}%
\providecommand \bibnamefont  [1]{#1}%
\providecommand \bibfnamefont [1]{#1}%
\providecommand \citenamefont [1]{#1}%
\providecommand \href@noop [0]{\@secondoftwo}%
\providecommand \href [0]{\begingroup \@sanitize@url \@href}%
\providecommand \@href[1]{\@@startlink{#1}\@@href}%
\providecommand \@@href[1]{\endgroup#1\@@endlink}%
\providecommand \@sanitize@url [0]{\catcode `\\12\catcode `\$12\catcode
  `\&12\catcode `\#12\catcode `\^12\catcode `\_12\catcode `\%12\relax}%
\providecommand \@@startlink[1]{}%
\providecommand \@@endlink[0]{}%
\providecommand \url  [0]{\begingroup\@sanitize@url \@url }%
\providecommand \@url [1]{\endgroup\@href {#1}{\urlprefix }}%
\providecommand \urlprefix  [0]{URL }%
\providecommand \Eprint [0]{\href }%
\providecommand \doibase [0]{https://doi.org/}%
\providecommand \selectlanguage [0]{\@gobble}%
\providecommand \bibinfo  [0]{\@secondoftwo}%
\providecommand \bibfield  [0]{\@secondoftwo}%
\providecommand \translation [1]{[#1]}%
\providecommand \BibitemOpen [0]{}%
\providecommand \bibitemStop [0]{}%
\providecommand \bibitemNoStop [0]{.\EOS\space}%
\providecommand \EOS [0]{\spacefactor3000\relax}%
\providecommand \BibitemShut  [1]{\csname bibitem#1\endcsname}%
\let\auto@bib@innerbib\@empty
\bibitem [{\citenamefont {Abbott}\ \emph {et~al.}(2021)\citenamefont {Abbott}
  \emph {et~al.}}]{LIGOScientific:2021sio}%
  \BibitemOpen
  \bibfield  {author} {\bibinfo {author} {\bibfnamefont {R.}~\bibnamefont
  {Abbott}} \emph {et~al.} (\bibinfo {collaboration} {LIGO Scientific, VIRGO,
  KAGRA}),\ }\href@noop {} {\  (\bibinfo {year} {2021})},\ \Eprint
  {https://arxiv.org/abs/2112.06861} {arXiv:2112.06861 [gr-qc]} \BibitemShut
  {NoStop}%
\bibitem [{\citenamefont {Will}(2018)}]{Will:2018bme}%
  \BibitemOpen
  \bibfield  {author} {\bibinfo {author} {\bibfnamefont {C.~M.}\ \bibnamefont
  {Will}},\ }\href@noop {} {\emph {\bibinfo {title} {{Theory and Experiment in
  Gravitational Physics}}}}\ (\bibinfo  {publisher} {Cambridge University
  Press},\ \bibinfo {year} {2018})\BibitemShut {NoStop}%
\bibitem [{\citenamefont {Flanagan}\ and\ \citenamefont
  {Hughes}(1998)}]{Flanagan:1997sx}%
  \BibitemOpen
  \bibfield  {author} {\bibinfo {author} {\bibfnamefont {E.~E.}\ \bibnamefont
  {Flanagan}}\ and\ \bibinfo {author} {\bibfnamefont {S.~A.}\ \bibnamefont
  {Hughes}},\ }\href {https://doi.org/10.1103/PhysRevD.57.4535} {\bibfield
  {journal} {\bibinfo  {journal} {Phys. Rev. D}\ }\textbf {\bibinfo {volume}
  {57}},\ \bibinfo {pages} {4535} (\bibinfo {year} {1998})},\ \Eprint
  {https://arxiv.org/abs/gr-qc/9701039} {arXiv:gr-qc/9701039} \BibitemShut
  {NoStop}%
\bibitem [{\citenamefont {Clifton}\ \emph {et~al.}(2012)\citenamefont
  {Clifton}, \citenamefont {Ferreira}, \citenamefont {Padilla},\ and\
  \citenamefont {Skordis}}]{Clifton:2011jh}%
  \BibitemOpen
  \bibfield  {author} {\bibinfo {author} {\bibfnamefont {T.}~\bibnamefont
  {Clifton}}, \bibinfo {author} {\bibfnamefont {P.~G.}\ \bibnamefont
  {Ferreira}}, \bibinfo {author} {\bibfnamefont {A.}~\bibnamefont {Padilla}},\
  and\ \bibinfo {author} {\bibfnamefont {C.}~\bibnamefont {Skordis}},\ }\href
  {https://doi.org/10.1016/j.physrep.2012.01.001} {\bibfield  {journal}
  {\bibinfo  {journal} {Phys. Rept.}\ }\textbf {\bibinfo {volume} {513}},\
  \bibinfo {pages} {1} (\bibinfo {year} {2012})},\ \Eprint
  {https://arxiv.org/abs/1106.2476} {arXiv:1106.2476 [astro-ph.CO]}
  \BibitemShut {NoStop}%
\bibitem [{\citenamefont {Delsate}\ \emph {et~al.}(2015)\citenamefont
  {Delsate}, \citenamefont {Hilditch},\ and\ \citenamefont
  {Witek}}]{Delsate:2014hba}%
  \BibitemOpen
  \bibfield  {author} {\bibinfo {author} {\bibfnamefont {T.}~\bibnamefont
  {Delsate}}, \bibinfo {author} {\bibfnamefont {D.}~\bibnamefont {Hilditch}},\
  and\ \bibinfo {author} {\bibfnamefont {H.}~\bibnamefont {Witek}},\ }\href
  {https://doi.org/10.1103/PhysRevD.91.024027} {\bibfield  {journal} {\bibinfo
  {journal} {Phys. Rev. D}\ }\textbf {\bibinfo {volume} {91}},\ \bibinfo
  {pages} {024027} (\bibinfo {year} {2015})},\ \Eprint
  {https://arxiv.org/abs/1407.6727} {arXiv:1407.6727 [gr-qc]} \BibitemShut
  {NoStop}%
\bibitem [{\citenamefont {Papallo}\ and\ \citenamefont
  {Reall}(2017)}]{Papallo:2017qvl}%
  \BibitemOpen
  \bibfield  {author} {\bibinfo {author} {\bibfnamefont {G.}~\bibnamefont
  {Papallo}}\ and\ \bibinfo {author} {\bibfnamefont {H.~S.}\ \bibnamefont
  {Reall}},\ }\href {https://doi.org/10.1103/PhysRevD.96.044019} {\bibfield
  {journal} {\bibinfo  {journal} {Phys. Rev. D}\ }\textbf {\bibinfo {volume}
  {96}},\ \bibinfo {pages} {044019} (\bibinfo {year} {2017})},\ \Eprint
  {https://arxiv.org/abs/1705.04370} {arXiv:1705.04370 [gr-qc]} \BibitemShut
  {NoStop}%
\bibitem [{\citenamefont {Ripley}\ and\ \citenamefont
  {Pretorius}(2019)}]{Ripley:2019hxt}%
  \BibitemOpen
  \bibfield  {author} {\bibinfo {author} {\bibfnamefont {J.~L.}\ \bibnamefont
  {Ripley}}\ and\ \bibinfo {author} {\bibfnamefont {F.}~\bibnamefont
  {Pretorius}},\ }\href {https://doi.org/10.1103/PhysRevD.99.084014} {\bibfield
   {journal} {\bibinfo  {journal} {Phys. Rev. D}\ }\textbf {\bibinfo {volume}
  {99}},\ \bibinfo {pages} {084014} (\bibinfo {year} {2019})},\ \Eprint
  {https://arxiv.org/abs/1902.01468} {arXiv:1902.01468 [gr-qc]} \BibitemShut
  {NoStop}%
\bibitem [{\citenamefont {Bernard}\ \emph {et~al.}(2019)\citenamefont
  {Bernard}, \citenamefont {Lehner},\ and\ \citenamefont
  {Luna}}]{Bernard:2019fjb}%
  \BibitemOpen
  \bibfield  {author} {\bibinfo {author} {\bibfnamefont {L.}~\bibnamefont
  {Bernard}}, \bibinfo {author} {\bibfnamefont {L.}~\bibnamefont {Lehner}},\
  and\ \bibinfo {author} {\bibfnamefont {R.}~\bibnamefont {Luna}},\ }\href
  {https://doi.org/10.1103/PhysRevD.100.024011} {\bibfield  {journal} {\bibinfo
   {journal} {Phys. Rev. D}\ }\textbf {\bibinfo {volume} {100}},\ \bibinfo
  {pages} {024011} (\bibinfo {year} {2019})},\ \Eprint
  {https://arxiv.org/abs/1904.12866} {arXiv:1904.12866 [gr-qc]} \BibitemShut
  {NoStop}%
\bibitem [{\citenamefont {Okounkova}\ \emph {et~al.}(2019)\citenamefont
  {Okounkova}, \citenamefont {Stein}, \citenamefont {Scheel},\ and\
  \citenamefont {Teukolsky}}]{Okounkova:2019dfo}%
  \BibitemOpen
  \bibfield  {author} {\bibinfo {author} {\bibfnamefont {M.}~\bibnamefont
  {Okounkova}}, \bibinfo {author} {\bibfnamefont {L.~C.}\ \bibnamefont
  {Stein}}, \bibinfo {author} {\bibfnamefont {M.~A.}\ \bibnamefont {Scheel}},\
  and\ \bibinfo {author} {\bibfnamefont {S.~A.}\ \bibnamefont {Teukolsky}},\
  }\href {https://doi.org/10.1103/PhysRevD.100.104026} {\bibfield  {journal}
  {\bibinfo  {journal} {Phys. Rev. D}\ }\textbf {\bibinfo {volume} {100}},\
  \bibinfo {pages} {104026} (\bibinfo {year} {2019})},\ \Eprint
  {https://arxiv.org/abs/1906.08789} {arXiv:1906.08789 [gr-qc]} \BibitemShut
  {NoStop}%
\bibitem [{\citenamefont {Figueras}\ and\ \citenamefont
  {Fran\c{c}a}(2020)}]{Figueras:2020dzx}%
  \BibitemOpen
  \bibfield  {author} {\bibinfo {author} {\bibfnamefont {P.}~\bibnamefont
  {Figueras}}\ and\ \bibinfo {author} {\bibfnamefont {T.}~\bibnamefont
  {Fran\c{c}a}},\ }\href {https://doi.org/10.1088/1361-6382/abb693} {\bibfield
  {journal} {\bibinfo  {journal} {Class. Quant. Grav.}\ }\textbf {\bibinfo
  {volume} {37}},\ \bibinfo {pages} {225009} (\bibinfo {year} {2020})},\
  \Eprint {https://arxiv.org/abs/2006.09414} {arXiv:2006.09414 [gr-qc]}
  \BibitemShut {NoStop}%
\bibitem [{\citenamefont {Gerhardinger}\ \emph {et~al.}(2022)\citenamefont
  {Gerhardinger}, \citenamefont {Giblin}, \citenamefont {Tolley},\ and\
  \citenamefont {Trodden}}]{Gerhardinger:2022bcw}%
  \BibitemOpen
  \bibfield  {author} {\bibinfo {author} {\bibfnamefont {M.}~\bibnamefont
  {Gerhardinger}}, \bibinfo {author} {\bibfnamefont {J.~T.}\ \bibnamefont
  {Giblin}, \bibfnamefont {Jr.}}, \bibinfo {author} {\bibfnamefont {A.~J.}\
  \bibnamefont {Tolley}},\ and\ \bibinfo {author} {\bibfnamefont
  {M.}~\bibnamefont {Trodden}},\ }\href
  {https://doi.org/10.1103/PhysRevD.106.043522} {\bibfield  {journal} {\bibinfo
   {journal} {Phys. Rev. D}\ }\textbf {\bibinfo {volume} {106}},\ \bibinfo
  {pages} {043522} (\bibinfo {year} {2022})},\ \Eprint
  {https://arxiv.org/abs/2205.05697} {arXiv:2205.05697 [hep-th]} \BibitemShut
  {NoStop}%
\bibitem [{198(1989)}]{1989iii}%
  \BibitemOpen
  in\ \href {https://doi.org/https://doi.org/10.1016/S0079-8169(08)62299-0}
  {\emph {\bibinfo {booktitle} {Initial-Boundary Value Problems and the
  Navier-Stokes Equations}}},\ \bibinfo {series} {Pure and Applied
  Mathematics}, Vol.\ \bibinfo {volume} {136},\ \bibinfo {editor} {edited by\
  \bibinfo {editor} {\bibfnamefont {H.-O.}\ \bibnamefont {Kreiss}}\ and\
  \bibinfo {editor} {\bibfnamefont {J.}~\bibnamefont {Lorenz}}}\ (\bibinfo
  {publisher} {Elsevier},\ \bibinfo {year} {1989})\ p.\ \bibinfo {pages}
  {iii}\BibitemShut {NoStop}%
\bibitem [{\citenamefont {Burgess}(2007)}]{Burgess:2007pt}%
  \BibitemOpen
  \bibfield  {author} {\bibinfo {author} {\bibfnamefont {C.~P.}\ \bibnamefont
  {Burgess}},\ }\href {https://doi.org/10.1146/annurev.nucl.56.080805.140508}
  {\bibfield  {journal} {\bibinfo  {journal} {Ann. Rev. Nucl. Part. Sci.}\
  }\textbf {\bibinfo {volume} {57}},\ \bibinfo {pages} {329} (\bibinfo {year}
  {2007})},\ \Eprint {https://arxiv.org/abs/hep-th/0701053}
  {arXiv:hep-th/0701053} \BibitemShut {NoStop}%
\bibitem [{\citenamefont {{Israel}}\ and\ \citenamefont
  {{Stewart}}(1979)}]{1979AnPhy.118..341I}%
  \BibitemOpen
  \bibfield  {author} {\bibinfo {author} {\bibfnamefont {W.}~\bibnamefont
  {{Israel}}}\ and\ \bibinfo {author} {\bibfnamefont {J.~M.}\ \bibnamefont
  {{Stewart}}},\ }\href {https://doi.org/10.1016/0003-4916(79)90130-1}
  {\bibfield  {journal} {\bibinfo  {journal} {Annals of Physics}\ }\textbf
  {\bibinfo {volume} {118}},\ \bibinfo {pages} {341} (\bibinfo {year}
  {1979})}\BibitemShut {NoStop}%
\bibitem [{\citenamefont {Cayuso}\ \emph {et~al.}(2017)\citenamefont {Cayuso},
  \citenamefont {Ortiz},\ and\ \citenamefont {Lehner}}]{Cayuso:2017iqc}%
  \BibitemOpen
  \bibfield  {author} {\bibinfo {author} {\bibfnamefont {J.}~\bibnamefont
  {Cayuso}}, \bibinfo {author} {\bibfnamefont {N.}~\bibnamefont {Ortiz}},\ and\
  \bibinfo {author} {\bibfnamefont {L.}~\bibnamefont {Lehner}},\ }\href
  {https://doi.org/10.1103/PhysRevD.96.084043} {\bibfield  {journal} {\bibinfo
  {journal} {Phys. Rev. D}\ }\textbf {\bibinfo {volume} {96}},\ \bibinfo
  {pages} {084043} (\bibinfo {year} {2017})},\ \Eprint
  {https://arxiv.org/abs/1706.07421} {arXiv:1706.07421 [gr-qc]} \BibitemShut
  {NoStop}%
\bibitem [{\citenamefont {Allwright}\ and\ \citenamefont
  {Lehner}(2019)}]{Allwright:2018rut}%
  \BibitemOpen
  \bibfield  {author} {\bibinfo {author} {\bibfnamefont {G.}~\bibnamefont
  {Allwright}}\ and\ \bibinfo {author} {\bibfnamefont {L.}~\bibnamefont
  {Lehner}},\ }\href {https://doi.org/10.1088/1361-6382/ab0ee1} {\bibfield
  {journal} {\bibinfo  {journal} {Class. Quant. Grav.}\ }\textbf {\bibinfo
  {volume} {36}},\ \bibinfo {pages} {084001} (\bibinfo {year} {2019})},\
  \Eprint {https://arxiv.org/abs/1808.07897} {arXiv:1808.07897 [gr-qc]}
  \BibitemShut {NoStop}%
\bibitem [{\citenamefont {Cayuso}\ and\ \citenamefont
  {Lehner}(2020)}]{Cayuso:2020lca}%
  \BibitemOpen
  \bibfield  {author} {\bibinfo {author} {\bibfnamefont {R.}~\bibnamefont
  {Cayuso}}\ and\ \bibinfo {author} {\bibfnamefont {L.}~\bibnamefont
  {Lehner}},\ }\href {https://doi.org/10.1103/PhysRevD.102.084008} {\bibfield
  {journal} {\bibinfo  {journal} {Phys. Rev. D}\ }\textbf {\bibinfo {volume}
  {102}},\ \bibinfo {pages} {084008} (\bibinfo {year} {2020})},\ \Eprint
  {https://arxiv.org/abs/2005.13720} {arXiv:2005.13720 [gr-qc]} \BibitemShut
  {NoStop}%
\bibitem [{\citenamefont {Franchini}\ \emph {et~al.}(2022)\citenamefont
  {Franchini}, \citenamefont {Bezares}, \citenamefont {Barausse},\ and\
  \citenamefont {Lehner}}]{Franchini:2022ukz}%
  \BibitemOpen
  \bibfield  {author} {\bibinfo {author} {\bibfnamefont {N.}~\bibnamefont
  {Franchini}}, \bibinfo {author} {\bibfnamefont {M.}~\bibnamefont {Bezares}},
  \bibinfo {author} {\bibfnamefont {E.}~\bibnamefont {Barausse}},\ and\
  \bibinfo {author} {\bibfnamefont {L.}~\bibnamefont {Lehner}},\ }\href
  {https://doi.org/10.1103/PhysRevD.106.064061} {\bibfield  {journal} {\bibinfo
   {journal} {Phys. Rev. D}\ }\textbf {\bibinfo {volume} {106}},\ \bibinfo
  {pages} {064061} (\bibinfo {year} {2022})},\ \Eprint
  {https://arxiv.org/abs/2206.00014} {arXiv:2206.00014 [gr-qc]} \BibitemShut
  {NoStop}%
\bibitem [{\citenamefont {Endlich}\ \emph {et~al.}(2017)\citenamefont
  {Endlich}, \citenamefont {Gorbenko}, \citenamefont {Huang},\ and\
  \citenamefont {Senatore}}]{Endlich:2017tqa}%
  \BibitemOpen
  \bibfield  {author} {\bibinfo {author} {\bibfnamefont {S.}~\bibnamefont
  {Endlich}}, \bibinfo {author} {\bibfnamefont {V.}~\bibnamefont {Gorbenko}},
  \bibinfo {author} {\bibfnamefont {J.}~\bibnamefont {Huang}},\ and\ \bibinfo
  {author} {\bibfnamefont {L.}~\bibnamefont {Senatore}},\ }\href
  {https://doi.org/10.1007/JHEP09(2017)122} {\bibfield  {journal} {\bibinfo
  {journal} {JHEP}\ }\textbf {\bibinfo {volume} {09}},\ \bibinfo {pages}
  {122}},\ \Eprint {https://arxiv.org/abs/1704.01590} {arXiv:1704.01590
  [gr-qc]} \BibitemShut {NoStop}%
\bibitem [{\citenamefont {Sarbach}\ and\ \citenamefont
  {Tiglio}(2012)}]{Sarbach:2012pr}%
  \BibitemOpen
  \bibfield  {author} {\bibinfo {author} {\bibfnamefont {O.}~\bibnamefont
  {Sarbach}}\ and\ \bibinfo {author} {\bibfnamefont {M.}~\bibnamefont
  {Tiglio}},\ }\href {https://doi.org/10.12942/lrr-2012-9} {\bibfield
  {journal} {\bibinfo  {journal} {Living Rev. Rel.}\ }\textbf {\bibinfo
  {volume} {15}},\ \bibinfo {pages} {9} (\bibinfo {year} {2012})},\ \Eprint
  {https://arxiv.org/abs/1203.6443} {arXiv:1203.6443 [gr-qc]} \BibitemShut
  {NoStop}%
\bibitem [{\citenamefont {Okounkova}\ \emph {et~al.}(2020)\citenamefont
  {Okounkova}, \citenamefont {Stein}, \citenamefont {Moxon}, \citenamefont
  {Scheel},\ and\ \citenamefont {Teukolsky}}]{Okounkova:2019zjf}%
  \BibitemOpen
  \bibfield  {author} {\bibinfo {author} {\bibfnamefont {M.}~\bibnamefont
  {Okounkova}}, \bibinfo {author} {\bibfnamefont {L.~C.}\ \bibnamefont
  {Stein}}, \bibinfo {author} {\bibfnamefont {J.}~\bibnamefont {Moxon}},
  \bibinfo {author} {\bibfnamefont {M.~A.}\ \bibnamefont {Scheel}},\ and\
  \bibinfo {author} {\bibfnamefont {S.~A.}\ \bibnamefont {Teukolsky}},\ }\href
  {https://doi.org/10.1103/PhysRevD.101.104016} {\bibfield  {journal} {\bibinfo
   {journal} {Phys. Rev. D}\ }\textbf {\bibinfo {volume} {101}},\ \bibinfo
  {pages} {104016} (\bibinfo {year} {2020})},\ \Eprint
  {https://arxiv.org/abs/1911.02588} {arXiv:1911.02588 [gr-qc]} \BibitemShut
  {NoStop}%
\bibitem [{\citenamefont {Reall}\ and\ \citenamefont
  {Warnick}(2022)}]{Reall:2021ebq}%
  \BibitemOpen
  \bibfield  {author} {\bibinfo {author} {\bibfnamefont {H.~S.}\ \bibnamefont
  {Reall}}\ and\ \bibinfo {author} {\bibfnamefont {C.~M.}\ \bibnamefont
  {Warnick}},\ }\href {https://doi.org/10.1063/5.0075455} {\bibfield  {journal}
  {\bibinfo  {journal} {J. Math. Phys.}\ }\textbf {\bibinfo {volume} {63}},\
  \bibinfo {pages} {042901} (\bibinfo {year} {2022})},\ \Eprint
  {https://arxiv.org/abs/2105.12028} {arXiv:2105.12028 [hep-th]} \BibitemShut
  {NoStop}%
\bibitem [{\citenamefont {Brodbeck}\ \emph {et~al.}(1999)\citenamefont
  {Brodbeck}, \citenamefont {Frittelli}, \citenamefont {Hubner},\ and\
  \citenamefont {Reula}}]{Brodbeck:1998az}%
  \BibitemOpen
  \bibfield  {author} {\bibinfo {author} {\bibfnamefont {O.}~\bibnamefont
  {Brodbeck}}, \bibinfo {author} {\bibfnamefont {S.}~\bibnamefont {Frittelli}},
  \bibinfo {author} {\bibfnamefont {P.}~\bibnamefont {Hubner}},\ and\ \bibinfo
  {author} {\bibfnamefont {O.~A.}\ \bibnamefont {Reula}},\ }\href
  {https://doi.org/10.1063/1.532694} {\bibfield  {journal} {\bibinfo  {journal}
  {J. Math. Phys.}\ }\textbf {\bibinfo {volume} {40}},\ \bibinfo {pages} {909}
  (\bibinfo {year} {1999})},\ \Eprint {https://arxiv.org/abs/gr-qc/9809023}
  {arXiv:gr-qc/9809023} \BibitemShut {NoStop}%
\bibitem [{\citenamefont {Gundlach}\ \emph {et~al.}(2005)\citenamefont
  {Gundlach}, \citenamefont {Martin-Garcia}, \citenamefont {Calabrese},\ and\
  \citenamefont {Hinder}}]{Gundlach:2005eh}%
  \BibitemOpen
  \bibfield  {author} {\bibinfo {author} {\bibfnamefont {C.}~\bibnamefont
  {Gundlach}}, \bibinfo {author} {\bibfnamefont {J.~M.}\ \bibnamefont
  {Martin-Garcia}}, \bibinfo {author} {\bibfnamefont {G.}~\bibnamefont
  {Calabrese}},\ and\ \bibinfo {author} {\bibfnamefont {I.}~\bibnamefont
  {Hinder}},\ }\href {https://doi.org/10.1088/0264-9381/22/17/025} {\bibfield
  {journal} {\bibinfo  {journal} {Class. Quant. Grav.}\ }\textbf {\bibinfo
  {volume} {22}},\ \bibinfo {pages} {3767} (\bibinfo {year} {2005})},\ \Eprint
  {https://arxiv.org/abs/gr-qc/0504114} {arXiv:gr-qc/0504114} \BibitemShut
  {NoStop}%
\bibitem [{\citenamefont {Kovacs}(2021)}]{Kovacs:2021lgk}%
  \BibitemOpen
  \bibfield  {author} {\bibinfo {author} {\bibfnamefont {A.~D.}\ \bibnamefont
  {Kovacs}},\ }\Eprint {https://arxiv.org/abs/2103.06895} {arXiv:2103.06895
  [gr-qc]}  (\bibinfo {year} {2021})\BibitemShut {NoStop}%
\bibitem [{\citenamefont {Lichnerowicz}(1944)}]{Lichnerowicz:1944}%
  \BibitemOpen
  \bibfield  {author} {\bibinfo {author} {\bibfnamefont {A.}~\bibnamefont
  {Lichnerowicz}},\ }\href@noop {} {\bibfield  {journal} {\bibinfo  {journal}
  {J. Math. Pures et Appl.}\ }\textbf {\bibinfo {volume} {23}},\ \bibinfo
  {pages} {37} (\bibinfo {year} {1944})}\BibitemShut {NoStop}%
\bibitem [{\citenamefont {York}(1971)}]{York:1971hw}%
  \BibitemOpen
  \bibfield  {author} {\bibinfo {author} {\bibfnamefont {J.~W.}\ \bibnamefont
  {York}, \bibfnamefont {Jr.}},\ }\href
  {https://doi.org/10.1103/PhysRevLett.26.1656} {\bibfield  {journal} {\bibinfo
   {journal} {Phys. Rev. Lett.}\ }\textbf {\bibinfo {volume} {26}},\ \bibinfo
  {pages} {1656} (\bibinfo {year} {1971})}\BibitemShut {NoStop}%
\bibitem [{\citenamefont {York}(1972)}]{York:1972sj}%
  \BibitemOpen
  \bibfield  {author} {\bibinfo {author} {\bibfnamefont {J.~W.}\ \bibnamefont
  {York}, \bibfnamefont {Jr.}},\ }\href
  {https://doi.org/10.1103/PhysRevLett.28.1082} {\bibfield  {journal} {\bibinfo
   {journal} {Phys. Rev. Lett.}\ }\textbf {\bibinfo {volume} {28}},\ \bibinfo
  {pages} {1082} (\bibinfo {year} {1972})}\BibitemShut {NoStop}%
\bibitem [{\citenamefont {Bowen}\ and\ \citenamefont
  {York}(1980)}]{Bowen:1980yu}%
  \BibitemOpen
  \bibfield  {author} {\bibinfo {author} {\bibfnamefont {J.~M.}\ \bibnamefont
  {Bowen}}\ and\ \bibinfo {author} {\bibfnamefont {J.~W.}\ \bibnamefont {York},
  \bibfnamefont {Jr.}},\ }\href {https://doi.org/10.1103/PhysRevD.21.2047}
  {\bibfield  {journal} {\bibinfo  {journal} {Phys. Rev. D}\ }\textbf {\bibinfo
  {volume} {21}},\ \bibinfo {pages} {2047} (\bibinfo {year}
  {1980})}\BibitemShut {NoStop}%
\bibitem [{\citenamefont {Clough}\ \emph {et~al.}(2015)\citenamefont {Clough},
  \citenamefont {Figueras}, \citenamefont {Finkel}, \citenamefont {Kunesch},
  \citenamefont {Lim},\ and\ \citenamefont {Tunyasuvunakool}}]{Clough:2015sqa}%
  \BibitemOpen
  \bibfield  {author} {\bibinfo {author} {\bibfnamefont {K.}~\bibnamefont
  {Clough}}, \bibinfo {author} {\bibfnamefont {P.}~\bibnamefont {Figueras}},
  \bibinfo {author} {\bibfnamefont {H.}~\bibnamefont {Finkel}}, \bibinfo
  {author} {\bibfnamefont {M.}~\bibnamefont {Kunesch}}, \bibinfo {author}
  {\bibfnamefont {E.~A.}\ \bibnamefont {Lim}},\ and\ \bibinfo {author}
  {\bibfnamefont {S.}~\bibnamefont {Tunyasuvunakool}},\ }\href
  {https://doi.org/10.1088/0264-9381/32/24/245011} {\bibfield  {journal}
  {\bibinfo  {journal} {Class. Quant. Grav.}\ }\textbf {\bibinfo {volume}
  {32}},\ \bibinfo {pages} {245011} (\bibinfo {year} {2015})},\ \Eprint
  {https://arxiv.org/abs/1503.03436} {arXiv:1503.03436 [gr-qc]} \BibitemShut
  {NoStop}%
\bibitem [{\citenamefont {Andrade}\ \emph {et~al.}(2021)\citenamefont {Andrade}
  \emph {et~al.}}]{Andrade:2021rbd}%
  \BibitemOpen
  \bibfield  {author} {\bibinfo {author} {\bibfnamefont {T.}~\bibnamefont
  {Andrade}} \emph {et~al.},\ }\href {https://doi.org/10.21105/joss.03703}
  {\bibfield  {journal} {\bibinfo  {journal} {J. Open Source Softw.}\ }\textbf
  {\bibinfo {volume} {6}},\ \bibinfo {pages} {3703} (\bibinfo {year} {2021})},\
  \Eprint {https://arxiv.org/abs/2201.03458} {arXiv:2201.03458 [gr-qc]}
  \BibitemShut {NoStop}%
\bibitem [{\citenamefont {Alic}\ \emph {et~al.}(2012)\citenamefont {Alic},
  \citenamefont {Bona-Casas}, \citenamefont {Bona}, \citenamefont {Rezzolla},\
  and\ \citenamefont {Palenzuela}}]{Alic:2011gg}%
  \BibitemOpen
  \bibfield  {author} {\bibinfo {author} {\bibfnamefont {D.}~\bibnamefont
  {Alic}}, \bibinfo {author} {\bibfnamefont {C.}~\bibnamefont {Bona-Casas}},
  \bibinfo {author} {\bibfnamefont {C.}~\bibnamefont {Bona}}, \bibinfo {author}
  {\bibfnamefont {L.}~\bibnamefont {Rezzolla}},\ and\ \bibinfo {author}
  {\bibfnamefont {C.}~\bibnamefont {Palenzuela}},\ }\href
  {https://doi.org/10.1103/PhysRevD.85.064040} {\bibfield  {journal} {\bibinfo
  {journal} {Phys. Rev. D}\ }\textbf {\bibinfo {volume} {85}},\ \bibinfo
  {pages} {064040} (\bibinfo {year} {2012})},\ \Eprint
  {https://arxiv.org/abs/1106.2254} {arXiv:1106.2254 [gr-qc]} \BibitemShut
  {NoStop}%
\bibitem [{\citenamefont {Bernuzzi}\ and\ \citenamefont
  {Hilditch}(2010)}]{Bernuzzi:2009ex}%
  \BibitemOpen
  \bibfield  {author} {\bibinfo {author} {\bibfnamefont {S.}~\bibnamefont
  {Bernuzzi}}\ and\ \bibinfo {author} {\bibfnamefont {D.}~\bibnamefont
  {Hilditch}},\ }\href {https://doi.org/10.1103/PhysRevD.81.084003} {\bibfield
  {journal} {\bibinfo  {journal} {Phys. Rev. D}\ }\textbf {\bibinfo {volume}
  {81}},\ \bibinfo {pages} {084003} (\bibinfo {year} {2010})},\ \Eprint
  {https://arxiv.org/abs/0912.2920} {arXiv:0912.2920 [gr-qc]} \BibitemShut
  {NoStop}%
\bibitem [{\citenamefont {Calabrese}\ \emph {et~al.}(2004)\citenamefont
  {Calabrese}, \citenamefont {Lehner}, \citenamefont {Reula}, \citenamefont
  {Sarbach},\ and\ \citenamefont {Tiglio}}]{Calabrese:2003vx}%
  \BibitemOpen
  \bibfield  {author} {\bibinfo {author} {\bibfnamefont {G.}~\bibnamefont
  {Calabrese}}, \bibinfo {author} {\bibfnamefont {L.}~\bibnamefont {Lehner}},
  \bibinfo {author} {\bibfnamefont {O.}~\bibnamefont {Reula}}, \bibinfo
  {author} {\bibfnamefont {O.}~\bibnamefont {Sarbach}},\ and\ \bibinfo {author}
  {\bibfnamefont {M.}~\bibnamefont {Tiglio}},\ }\href
  {https://doi.org/10.1088/0264-9381/21/24/004} {\bibfield  {journal} {\bibinfo
   {journal} {Class. Quant. Grav.}\ }\textbf {\bibinfo {volume} {21}},\
  \bibinfo {pages} {5735} (\bibinfo {year} {2004})},\ \Eprint
  {https://arxiv.org/abs/gr-qc/0308007} {arXiv:gr-qc/0308007} \BibitemShut
  {NoStop}%
\bibitem [{\citenamefont {Alic}\ \emph {et~al.}(2013)\citenamefont {Alic},
  \citenamefont {Kastaun},\ and\ \citenamefont {Rezzolla}}]{Alic:2013xsa}%
  \BibitemOpen
  \bibfield  {author} {\bibinfo {author} {\bibfnamefont {D.}~\bibnamefont
  {Alic}}, \bibinfo {author} {\bibfnamefont {W.}~\bibnamefont {Kastaun}},\ and\
  \bibinfo {author} {\bibfnamefont {L.}~\bibnamefont {Rezzolla}},\ }\href
  {https://doi.org/10.1103/PhysRevD.88.064049} {\bibfield  {journal} {\bibinfo
  {journal} {Phys. Rev. D}\ }\textbf {\bibinfo {volume} {88}},\ \bibinfo
  {pages} {064049} (\bibinfo {year} {2013})},\ \Eprint
  {https://arxiv.org/abs/1307.7391} {arXiv:1307.7391 [gr-qc]} \BibitemShut
  {NoStop}%
\bibitem [{\citenamefont {Schnetter}(2010)}]{Schnetter:2010cz}%
  \BibitemOpen
  \bibfield  {author} {\bibinfo {author} {\bibfnamefont {E.}~\bibnamefont
  {Schnetter}},\ }\href {https://doi.org/10.1088/0264-9381/27/16/167001}
  {\bibfield  {journal} {\bibinfo  {journal} {Class. Quant. Grav.}\ }\textbf
  {\bibinfo {volume} {27}},\ \bibinfo {pages} {167001} (\bibinfo {year}
  {2010})},\ \Eprint {https://arxiv.org/abs/1003.0859} {arXiv:1003.0859
  [gr-qc]} \BibitemShut {NoStop}%
\bibitem [{\citenamefont {de~Rham}\ \emph {et~al.}(2022)\citenamefont
  {de~Rham}, \citenamefont {Tolley},\ and\ \citenamefont
  {Zhang}}]{deRham:2021bll}%
  \BibitemOpen
  \bibfield  {author} {\bibinfo {author} {\bibfnamefont {C.}~\bibnamefont
  {de~Rham}}, \bibinfo {author} {\bibfnamefont {A.~J.}\ \bibnamefont
  {Tolley}},\ and\ \bibinfo {author} {\bibfnamefont {J.}~\bibnamefont
  {Zhang}},\ }\href {https://doi.org/10.1103/PhysRevLett.128.131102} {\bibfield
   {journal} {\bibinfo  {journal} {Phys. Rev. Lett.}\ }\textbf {\bibinfo
  {volume} {128}},\ \bibinfo {pages} {131102} (\bibinfo {year} {2022})},\
  \Eprint {https://arxiv.org/abs/2112.05054} {arXiv:2112.05054 [gr-qc]}
  \BibitemShut {NoStop}%
\bibitem [{\citenamefont {Sennett}\ \emph {et~al.}(2020)\citenamefont
  {Sennett}, \citenamefont {Brito}, \citenamefont {Buonanno}, \citenamefont
  {Gorbenko},\ and\ \citenamefont {Senatore}}]{Sennett:2019bpc}%
  \BibitemOpen
  \bibfield  {author} {\bibinfo {author} {\bibfnamefont {N.}~\bibnamefont
  {Sennett}}, \bibinfo {author} {\bibfnamefont {R.}~\bibnamefont {Brito}},
  \bibinfo {author} {\bibfnamefont {A.}~\bibnamefont {Buonanno}}, \bibinfo
  {author} {\bibfnamefont {V.}~\bibnamefont {Gorbenko}},\ and\ \bibinfo
  {author} {\bibfnamefont {L.}~\bibnamefont {Senatore}},\ }\href
  {https://doi.org/10.1103/PhysRevD.102.044056} {\bibfield  {journal} {\bibinfo
   {journal} {Phys. Rev. D}\ }\textbf {\bibinfo {volume} {102}},\ \bibinfo
  {pages} {044056} (\bibinfo {year} {2020})},\ \Eprint
  {https://arxiv.org/abs/1912.09917} {arXiv:1912.09917 [gr-qc]} \BibitemShut
  {NoStop}%
\bibitem [{\citenamefont {Silva}\ \emph {et~al.}(2023)\citenamefont {Silva},
  \citenamefont {Ghosh},\ and\ \citenamefont {Buonanno}}]{Silva:2022srr}%
  \BibitemOpen
  \bibfield  {author} {\bibinfo {author} {\bibfnamefont {H.~O.}\ \bibnamefont
  {Silva}}, \bibinfo {author} {\bibfnamefont {A.}~\bibnamefont {Ghosh}},\ and\
  \bibinfo {author} {\bibfnamefont {A.}~\bibnamefont {Buonanno}},\ }\href
  {https://doi.org/10.1103/PhysRevD.107.044030} {\bibfield  {journal} {\bibinfo
   {journal} {Phys. Rev. D}\ }\textbf {\bibinfo {volume} {107}},\ \bibinfo
  {pages} {044030} (\bibinfo {year} {2023})},\ \Eprint
  {https://arxiv.org/abs/2205.05132} {arXiv:2205.05132 [gr-qc]} \BibitemShut
  {NoStop}%
\bibitem [{\citenamefont {Dideron}\ \emph {et~al.}(2022)\citenamefont
  {Dideron}, \citenamefont {Mukherjee},\ and\ \citenamefont
  {Lehner}}]{Dideron:2022tap}%
  \BibitemOpen
  \bibfield  {author} {\bibinfo {author} {\bibfnamefont {G.}~\bibnamefont
  {Dideron}}, \bibinfo {author} {\bibfnamefont {S.}~\bibnamefont {Mukherjee}},\
  and\ \bibinfo {author} {\bibfnamefont {L.}~\bibnamefont {Lehner}},\ }\Eprint
  {https://arxiv.org/abs/2209.14321} {arXiv:2209.14321 [gr-qc]}  (\bibinfo
  {year} {2022})\BibitemShut {NoStop}%
\bibitem [{\citenamefont {Cardoso}\ \emph {et~al.}(2018)\citenamefont
  {Cardoso}, \citenamefont {Kimura}, \citenamefont {Maselli},\ and\
  \citenamefont {Senatore}}]{Cardoso:2018ptl}%
  \BibitemOpen
  \bibfield  {author} {\bibinfo {author} {\bibfnamefont {V.}~\bibnamefont
  {Cardoso}}, \bibinfo {author} {\bibfnamefont {M.}~\bibnamefont {Kimura}},
  \bibinfo {author} {\bibfnamefont {A.}~\bibnamefont {Maselli}},\ and\ \bibinfo
  {author} {\bibfnamefont {L.}~\bibnamefont {Senatore}},\ }\href
  {https://doi.org/10.1103/PhysRevLett.121.251105} {\bibfield  {journal}
  {\bibinfo  {journal} {Phys. Rev. Lett.}\ }\textbf {\bibinfo {volume} {121}},\
  \bibinfo {pages} {251105} (\bibinfo {year} {2018})},\ \Eprint
  {https://arxiv.org/abs/1808.08962} {arXiv:1808.08962 [gr-qc]} \BibitemShut
  {NoStop}%
\bibitem [{\citenamefont {Flanagan}\ and\ \citenamefont
  {Hinderer}(2008)}]{Flanagan:2007ix}%
  \BibitemOpen
  \bibfield  {author} {\bibinfo {author} {\bibfnamefont {E.~E.}\ \bibnamefont
  {Flanagan}}\ and\ \bibinfo {author} {\bibfnamefont {T.}~\bibnamefont
  {Hinderer}},\ }\href {https://doi.org/10.1103/PhysRevD.77.021502} {\bibfield
  {journal} {\bibinfo  {journal} {Phys. Rev. D}\ }\textbf {\bibinfo {volume}
  {77}},\ \bibinfo {pages} {021502} (\bibinfo {year} {2008})},\ \Eprint
  {https://arxiv.org/abs/0709.1915} {arXiv:0709.1915 [astro-ph]} \BibitemShut
  {NoStop}%
\bibitem [{\citenamefont {Porto}(2016)}]{Porto:2016zng}%
  \BibitemOpen
  \bibfield  {author} {\bibinfo {author} {\bibfnamefont {R.~A.}\ \bibnamefont
  {Porto}},\ }\href {https://doi.org/10.1002/prop.201600064} {\bibfield
  {journal} {\bibinfo  {journal} {Fortsch. Phys.}\ }\textbf {\bibinfo {volume}
  {64}},\ \bibinfo {pages} {723} (\bibinfo {year} {2016})},\ \Eprint
  {https://arxiv.org/abs/1606.08895} {arXiv:1606.08895 [gr-qc]} \BibitemShut
  {NoStop}%
\bibitem [{\citenamefont {Corman}\ \emph {et~al.}(2023)\citenamefont {Corman},
  \citenamefont {Ripley},\ and\ \citenamefont {East}}]{Corman:2022xqg}%
  \BibitemOpen
  \bibfield  {author} {\bibinfo {author} {\bibfnamefont {M.}~\bibnamefont
  {Corman}}, \bibinfo {author} {\bibfnamefont {J.~L.}\ \bibnamefont {Ripley}},\
  and\ \bibinfo {author} {\bibfnamefont {W.~E.}\ \bibnamefont {East}},\ }\href
  {https://doi.org/10.1103/PhysRevD.107.024014} {\bibfield  {journal} {\bibinfo
   {journal} {Phys. Rev. D}\ }\textbf {\bibinfo {volume} {107}},\ \bibinfo
  {pages} {024014} (\bibinfo {year} {2023})},\ \Eprint
  {https://arxiv.org/abs/2210.09235} {arXiv:2210.09235 [gr-qc]} \BibitemShut
  {NoStop}%
\bibitem [{\citenamefont {Cano}\ \emph {et~al.}(2020)\citenamefont {Cano},
  \citenamefont {Fransen},\ and\ \citenamefont {Hertog}}]{Cano:2020cao}%
  \BibitemOpen
  \bibfield  {author} {\bibinfo {author} {\bibfnamefont {P.~A.}\ \bibnamefont
  {Cano}}, \bibinfo {author} {\bibfnamefont {K.}~\bibnamefont {Fransen}},\ and\
  \bibinfo {author} {\bibfnamefont {T.}~\bibnamefont {Hertog}},\ }\href
  {https://doi.org/10.1103/PhysRevD.102.044047} {\bibfield  {journal} {\bibinfo
   {journal} {Phys. Rev. D}\ }\textbf {\bibinfo {volume} {102}},\ \bibinfo
  {pages} {044047} (\bibinfo {year} {2020})},\ \Eprint
  {https://arxiv.org/abs/2005.03671} {arXiv:2005.03671 [gr-qc]} \BibitemShut
  {NoStop}%
\bibitem [{\citenamefont {Hollands}\ \emph {et~al.}(2022)\citenamefont
  {Hollands}, \citenamefont {Ishibashi},\ and\ \citenamefont
  {Reall}}]{Hollands:2022ajj}%
  \BibitemOpen
  \bibfield  {author} {\bibinfo {author} {\bibfnamefont {S.}~\bibnamefont
  {Hollands}}, \bibinfo {author} {\bibfnamefont {A.}~\bibnamefont
  {Ishibashi}},\ and\ \bibinfo {author} {\bibfnamefont {H.~S.}\ \bibnamefont
  {Reall}},\ }\Eprint {https://arxiv.org/abs/2212.06554} {arXiv:2212.06554
  [gr-qc]}  (\bibinfo {year} {2022})\BibitemShut {NoStop}%
\bibitem [{\citenamefont {Buonanno}\ \emph {et~al.}(2008)\citenamefont
  {Buonanno}, \citenamefont {Kidder},\ and\ \citenamefont
  {Lehner}}]{Buonanno:2007sv}%
  \BibitemOpen
  \bibfield  {author} {\bibinfo {author} {\bibfnamefont {A.}~\bibnamefont
  {Buonanno}}, \bibinfo {author} {\bibfnamefont {L.~E.}\ \bibnamefont
  {Kidder}},\ and\ \bibinfo {author} {\bibfnamefont {L.}~\bibnamefont
  {Lehner}},\ }\href {https://doi.org/10.1103/PhysRevD.77.026004} {\bibfield
  {journal} {\bibinfo  {journal} {Phys. Rev. D}\ }\textbf {\bibinfo {volume}
  {77}},\ \bibinfo {pages} {026004} (\bibinfo {year} {2008})},\ \Eprint
  {https://arxiv.org/abs/0709.3839} {arXiv:0709.3839 [astro-ph]} \BibitemShut
  {NoStop}%
\bibitem [{\citenamefont {Arest\'e~Sal\'o}\ \emph {et~al.}(2022)\citenamefont
  {Arest\'e~Sal\'o}, \citenamefont {Clough},\ and\ \citenamefont
  {Figueras}}]{AresteSalo:2022hua}%
  \BibitemOpen
  \bibfield  {author} {\bibinfo {author} {\bibfnamefont {L.}~\bibnamefont
  {Arest\'e~Sal\'o}}, \bibinfo {author} {\bibfnamefont {K.}~\bibnamefont
  {Clough}},\ and\ \bibinfo {author} {\bibfnamefont {P.}~\bibnamefont
  {Figueras}},\ }\href {https://doi.org/10.1103/PhysRevLett.129.261104}
  {\bibfield  {journal} {\bibinfo  {journal} {Phys. Rev. Lett.}\ }\textbf
  {\bibinfo {volume} {129}},\ \bibinfo {pages} {261104} (\bibinfo {year}
  {2022})},\ \Eprint {https://arxiv.org/abs/2208.14470} {arXiv:2208.14470
  [gr-qc]} \BibitemShut {NoStop}%
\bibitem [{\citenamefont {Bezares}\ \emph {et~al.}(2022)\citenamefont
  {Bezares}, \citenamefont {Aguilera-Miret}, \citenamefont {ter Haar},
  \citenamefont {Crisostomi}, \citenamefont {Palenzuela},\ and\ \citenamefont
  {Barausse}}]{Bezares:2021dma}%
  \BibitemOpen
  \bibfield  {author} {\bibinfo {author} {\bibfnamefont {M.}~\bibnamefont
  {Bezares}}, \bibinfo {author} {\bibfnamefont {R.}~\bibnamefont
  {Aguilera-Miret}}, \bibinfo {author} {\bibfnamefont {L.}~\bibnamefont {ter
  Haar}}, \bibinfo {author} {\bibfnamefont {M.}~\bibnamefont {Crisostomi}},
  \bibinfo {author} {\bibfnamefont {C.}~\bibnamefont {Palenzuela}},\ and\
  \bibinfo {author} {\bibfnamefont {E.}~\bibnamefont {Barausse}},\ }\href
  {https://doi.org/10.1103/PhysRevLett.128.091103} {\bibfield  {journal}
  {\bibinfo  {journal} {Phys. Rev. Lett.}\ }\textbf {\bibinfo {volume} {128}},\
  \bibinfo {pages} {091103} (\bibinfo {year} {2022})},\ \Eprint
  {https://arxiv.org/abs/2107.05648} {arXiv:2107.05648 [gr-qc]} \BibitemShut
  {NoStop}%
\bibitem [{\citenamefont {Radia}\ \emph {et~al.}(2022)\citenamefont {Radia},
  \citenamefont {Sperhake}, \citenamefont {Drew}, \citenamefont {Clough},
  \citenamefont {Figueras}, \citenamefont {Lim}, \citenamefont {Ripley},
  \citenamefont {Aurrekoetxea}, \citenamefont {Fran\c{c}a},\ and\ \citenamefont
  {Helfer}}]{Radia:2021smk}%
  \BibitemOpen
  \bibfield  {author} {\bibinfo {author} {\bibfnamefont {M.}~\bibnamefont
  {Radia}}, \bibinfo {author} {\bibfnamefont {U.}~\bibnamefont {Sperhake}},
  \bibinfo {author} {\bibfnamefont {A.}~\bibnamefont {Drew}}, \bibinfo {author}
  {\bibfnamefont {K.}~\bibnamefont {Clough}}, \bibinfo {author} {\bibfnamefont
  {P.}~\bibnamefont {Figueras}}, \bibinfo {author} {\bibfnamefont {E.~A.}\
  \bibnamefont {Lim}}, \bibinfo {author} {\bibfnamefont {J.~L.}\ \bibnamefont
  {Ripley}}, \bibinfo {author} {\bibfnamefont {J.~C.}\ \bibnamefont
  {Aurrekoetxea}}, \bibinfo {author} {\bibfnamefont {T.}~\bibnamefont
  {Fran\c{c}a}},\ and\ \bibinfo {author} {\bibfnamefont {T.}~\bibnamefont
  {Helfer}},\ }\href {https://doi.org/10.1088/1361-6382/ac6fa9} {\bibfield
  {journal} {\bibinfo  {journal} {Class. Quant. Grav.}\ }\textbf {\bibinfo
  {volume} {39}},\ \bibinfo {pages} {135006} (\bibinfo {year} {2022})},\
  \Eprint {https://arxiv.org/abs/2112.10567} {arXiv:2112.10567 [gr-qc]}
  \BibitemShut {NoStop}%
\bibitem [{\citenamefont {Fran\c{c}a}(2023)}]{tiago_thesis}%
  \BibitemOpen
  \bibfield  {author} {\bibinfo {author} {\bibfnamefont {T.}~\bibnamefont
  {Fran\c{c}a}},\ }\emph {\bibinfo {title} {{Binary Black Holes in Modified
  Gravity}}},\ \href@noop {} {Ph.D. thesis},\ \bibinfo  {school} {Queen Mary
  University of London} (\bibinfo {year} {2023})\BibitemShut {NoStop}%
\bibitem [{\citenamefont {Bezares}\ \emph {et~al.}(2021)\citenamefont
  {Bezares}, \citenamefont {ter Haar}, \citenamefont {Crisostomi},
  \citenamefont {Barausse},\ and\ \citenamefont
  {Palenzuela}}]{Bezares:2021yek}%
  \BibitemOpen
  \bibfield  {author} {\bibinfo {author} {\bibfnamefont {M.}~\bibnamefont
  {Bezares}}, \bibinfo {author} {\bibfnamefont {L.}~\bibnamefont {ter Haar}},
  \bibinfo {author} {\bibfnamefont {M.}~\bibnamefont {Crisostomi}}, \bibinfo
  {author} {\bibfnamefont {E.}~\bibnamefont {Barausse}},\ and\ \bibinfo
  {author} {\bibfnamefont {C.}~\bibnamefont {Palenzuela}},\ }\href
  {https://doi.org/10.1103/PhysRevD.104.044022} {\bibfield  {journal} {\bibinfo
   {journal} {Phys. Rev. D}\ }\textbf {\bibinfo {volume} {104}},\ \bibinfo
  {pages} {044022} (\bibinfo {year} {2021})},\ \Eprint
  {https://arxiv.org/abs/2105.13992} {arXiv:2105.13992 [gr-qc]} \BibitemShut
  {NoStop}%
\end{thebibliography}%


%

\end{document}